\documentclass[useAMS,usenatbib]{mn2e}

\usepackage[dvipdfmx]{graphicx}
\usepackage{subfigure}
\usepackage{float}
\usepackage{amsmath}
\usepackage{color}
\usepackage{ulem}

\RequirePackage[l2tabu, orthodox]{nag}

\newcommand{\apj}{ApJ}
\newcommand{\apjl}{ApJL}
\newcommand{\pasj}{PASJ}

\newcommand{\aj}{AJ} % Astronomical Journal
\newcommand{\mnras}{MNRAS}
\newcommand{\apjs}{ApJS}

\title[GMCs in a barred galaxy with star formation and thermal feedback]{GMC Evolution in a Barred Spiral Galaxy with Star Formation and Thermal Feedback}

\author[FUJIMOTO, BRYAN, TASKER, HABE \& SIMPSON]{Yusuke Fujimoto$^{1}$\thanks{E-mail: yfujimoto@frontier.hokudai.ac.jp}, Greg L. Bryan$^{2}$, Elizabeth J. Tasker$^{1}$, Asao Habe$^{1}$, and \newauthor Christine M. Simpson$^{3}$\\
$^{1}$Department of Physics, Faculty of Science, Hokkaido University, Kita 10 Nishi 8 Kita-ku, Sapporo 060-0810, Japan\\
$^{2}$Columbia University, Department of Astronomy, New York, NY 10025, USA\\
$^{3}$Heidelberger Institut f\"ur Theoretische Studien, Schloss-Wolfsbrunnenweg 35, 69118 Heidelberg, Germany}

\begin{document}

\pagerange{\pageref{firstpage}--\pageref{lastpage}} \pubyear{}
\maketitle
\label{firstpage}

\begin{abstract}
We explore the impact of star formation and thermal stellar feedback on the giant molecular cloud (GMC) population forming in a M83-type barred spiral galaxy. We compare three high-resolution simulations (1.5 pc cell size) with different star formation/feedback models: one with no star formation, one with star formation but no feedback, and one with star formation and thermal energy injection. We analyze the resulting population of clouds, finding that we can identify the same population of massive, virialized clouds and transient, low-surface density clouds found in our previous work (that did not include star formation or feedback).  Star formation and feedback can affect the mix of clouds we identify.  In particular, star formation alone simply converts dense cloud gas into stars with only a small change to the cloud populations, principally resulting in a slight decrease in the transient population.   Feedback, however, has a stronger impact: while it is not generally sufficient to entirely destroy the clouds, it does eject gas out of them, increasing the gas density in the inter-cloud region.  This decreases the number of massive clouds, but substantially increases the transient cloud population.  We also find that feedback tends to drive a net radial inflow of massive clouds, leading to an increase in the star formation rate in the bar region.  We examine a number of possible reasons for this and conclude that it is possible that the drag force from the enhanced intercloud density could be responsible.
\end{abstract}


%%%%%%%%%% Introduction %%%%%%%%%%
\section{Introduction}

Understanding the `where' and `how' of gas conversion into stars underpins many areas of astrophysics. In particular, it links the large, galaxy-scale evolution with the collapsing cores embedded within the giant molecular clouds. Key to this process is the evolution of the clouds themselves. Their properties, interactions and motion within the galaxy determine the rate and efficiency of the local star formation and from there, the global distribution. 

This global distribution is far from uniform. Observationally, there is the Kennicutt-Schmidt empirical power-law relation between the gas surface density and the surface density of the star formation rate \citep{Kennicutt1989, Kennicutt1998, Bigiel2008}. Yet closer inspection shows environmental differences between galaxy types and the nucleus, bar, spiral and inter-arm regions within a single galaxy \citep{Daddi2010, Leroy2013, Momose2010, Hirota2014, Muraoka2007}. These changes include systematic variations in the star formation efficiency ($\rm{SFE}=\Sigma_{\rm{SFR}}/\Sigma_{\rm{gas}}$) that suggest galactic structure plays a larger role than simply gathering gas so that it can form stars. Other factors beyond simply the local gas density are controlling the star formation rate. 

\citet{Koda2009} looked at GMCs in the spiral galaxy M51 and found that clouds assembled in the spiral arm to become giant molecular associations but broke apart into smaller clouds in the interarm region. This result was replicated in a simulation by \citet{Dobbs2006}. \citet{Meidt2013} and \citet{Colombo2014} also observed the same galaxy, finding that shearing flows and shocks from the spiral arms could stabilise the GMCs and prevent the formation of stars. This was supported by observations of the intermediate spiral galaxy, IC342 by \citet{Hirota2011}, who found that the GMCs that were forming stars were downstream of the spiral arm and tended to be more massive. This points to differences in the GMC population that control the star formation on local scales in these different regions. 

The impact of the global environment on cloud properties was investigated theoretically by \citet{Fujimoto2014a} (hereafter, Paper I). This work examined the GMC populations forming in the bar, spiral and quiescent outer disc regions of an M83-type galaxy with a grand design architecture. They found that while the typical GMC properties were uniform between the three environments, the cloud-cloud interaction rate was strongly dependent on the global structure. The elliptical gas motion in the bar and spiral potentials increased the interaction rate to create a larger number of both giant molecular associations from multiple collisions, and small, transient clouds found in tidal tails. The resultant three cloud types were clearly distinguished by their locations on the mass-radius scaling relation: {\it Type A} clouds were the most common cloud in all environments, with properties matching those typically observed in local galaxies with a median mass of $5\times10^5\  M_{\odot}$, radius $15\ \rm pc$ and velocity dispersion $6\ \rm km/s$ \citep{Solomon1987, Heyer2009, Roman-Duval2010, Muraoka2009}. {\it Type B} clouds were the giant molecular associations with radii above 30\,pc, created during multiple mergers, while {\it Type C} were unbound, transient clouds forming in the tidal tails, predominantly around the gravitationally dominating {\it Type B}s. 

The cloud interaction rate has previously been linked to the production of stars. In analytical calculations by \citet{Tan2000}, it was suggested that a frequent collision rate could drive the star formation to create the observed relation with gas surface density. The required rate was supported by simulations \citep{TaskerTan2009}, while observational evidence that such collisions could trigger star formation has been seen by \citet{Furukawa2009, Ohama2010, Fukui2014}. With this in mind, Paper I estimated the star formation rate based on the cloud interactions. This resulted in the highest star formation efficiency being in the bar region, which is not seen in observations \citep{Momose2010, Sorai2012}. However, simulations of individual cloud collisions suggested that high velocity interactions might be less productive than lower velocity encounters \citep{Takahira2014}. By incorporating this into the model, \citet{Fujimoto2014b} found that while the interactions were frequent in the bar region, they were too fast to produce a high yield of stars, resulting in a lower star formation efficiency in that region. While cloud interactions are not the only controlling factor in star formation, their importance --and therefore the importance of the cloud environment-- should not be underestimated.

While Paper I explored the environmental influences on GMC evolution, it did not include the internal processes of active star formation or stellar feedback. Stars emit UV radiation and stellar winds during their lifetime, with massive stars larger than 8\,M$_\odot$ dying in supernovae explosions. This energy is injected into the star's surrounding gaseous cradle, having a strong or even disruptive influence on the GMC. It is therefore another major player in determining star formation in the galaxy. Exactly how much of an impact this feedback has on GMCs and future star formation is strongly debated in the literature. For example, In simulations of a Milky Way-type galaxy, \citet{Tasker2015} found that thermal energy feedback could effectively suppress the star formation, but the surrounding GMC could survive to form more stars. This result was also seen by \citet{ShettyOstriker2008}, in two-dimensional simulations of the galactic disc, where the stars inject momentum into the gas. On the other hand, feedback was found to be a much more detrimental force for GMCs in simulations performed by \citet{Williamson2014}, creating a younger and less massive population than when no feedback is included. \citet{Dobbs2011} however, finds the profile slope of the GMCs is unaffected by thermal and momentum feedback for a star formation efficiency above 5\%, although its normalization does vary. The effect of different feedback mechanisms in simulations was tested by \citet{Hopkins2011, Hopkins2012}, who individually included the effects of supernovae, stellar winds, radiation pressure and HII photoionisation heating. They found that the star formation rates in the galaxy disc match the observed Kennicutt-Schmidt relation independent of the feedback used, providing it was effective at breaking up the densest gas.  \citet{Agertz2015} investigated the star formation-feedback cycle in cosmological galaxy simulations, finding that, in order to reproduce Milky Way properties, they required early momentum feedback and a large efficiency of star formation per dynamical time.

One way to understand these questions is to determine the source of ISM turbulence in the galactic disc. \citet{ElmegreenBurkert2010} found that turbulence could be driven by a combination of accretion, disk instabilities, and energetic feedback by young stars. \citet{Goldbaum2015, Goldbaum2016} explored the effects of gravitational instability and stellar feedback on turbulence in the disc, comparing a pure self-gravity model with one that included stellar feedback. They found that gravitational instabilities are likely to be the dominant source of turbulence, and that they can transport mass inwards, fueling star formation in the inner parts of galactic discs over cosmological time.

The interplay between the environmental and internal impact factors on the GMC has been less well studied. \citet{Dobbs2011} notes that the collection of gas in the galactic spiral arms to make giant molecular associations operates with and without feedback. However, there has not been a more detailed investigation as to whether internal feedback or external environment play the major role. In this paper, we explore these processes in a barred spiral galaxy simulation with resolutions down to 1.5\,pc, comparing the clouds formed in the galaxy disc. The two new simulations presented here have star formation without stellar feedback and a run which also includes localized thermal energy injection. We compare the results with those in Paper I, where no active star formation was included. Our simulations are based on M83, a nearby face-on galaxy, a target for the Atacama Large Millimeter Array (ALMA) in cycles 0, 1 and 2. This will ultimately allow our GMC populations to be compared with the high spatial resolution achieved in the ALMA observations.  The main analysis of our results is performed at $t = 200$\,Myr. This is earlier than in Paper I, which primarily focused on cloud properties at $t = 240$\,Myr. This change is due to the effectiveness of our feedback, which disperses the grand design spiral at later times, making the comparison with M83 difficult. In order to compare with future comparisons of M83 with instruments such as ALMA, we selected an earlier analysis time, when the grand design is still clearly present.

This paper is organized as follows: In Section~\ref{Numerical methods}, we present our model of the barred galaxy, along with the star formation, stellar feedback and cloud identification models. In Section~\ref{Results}, we describe our main findings, showing the effect of the stellar feedback on the ISM, GMC properties and star formation.

%%%%%%%%%% Numerical methods %%%%%%%%%%
\section{Numerical methods}
\label{Numerical methods}

%%%%%%%%%% subsection %%%%%%%%%%
\subsection{Simulation and initial conditions}

The simulations presented in this paper are of an isolated galaxy disc run using Enzo, a three-dimensional adaptive mesh refinement (AMR) hydrodynamics code \citep{Bryan2014}. The box size is 50\,kpc across, covered by a root grid of $128^3$ cells and eight levels of refinement, giving a limiting resolution (smallest cell size) of about 1.5\,pc. The cell was refined by a factor of two whenever the mass within the cell exceeded $1000\ \mathrm{M_{\odot}}$. We confirmed that this refinement criteria, even at our temperature floor, resolves the Jeans length  \citep{Truelove1997} up to a density of $10^4$ cm$^{-3}$, at which point we reach our maximum refinement.  Beyond this, we add an artificial pressure ($P \propto \rho^2$) to prevent unresolved collapse at the finest cell level so that the Jeans length is resolved by at least four cells \citep{Machacek2001}. 

The evolution of the gas was computed using a three-dimensional version of the ZEUS hydrodynamics algorithm \citep{StoneNorman1992}.  The gas was self-gravitating and allowed to cool radiatively down to 300\,K. The cooling rates were taken from the analytical expression of \citet{SarazinWhite1987} for solar metallicity down to $10^5$ K, and continued to 300 K with rates provided by \citet{RosenBregman1995}.

Our galaxy was modeled on the nearby barred spiral galaxy, M83. For a stellar potential, we used the model from \citet{Hirota2014}, who analysed the 2MASS $K$-band image of M83 to produce a description of the density in the axisymmetric bulge and disc and non-axisymmetric bar and spiral galactic regions. The non-axisymmetric stellar components consisted of $10^5$ fixed-motion star particles that rotate at the estimated pattern speed for M83, $54\ \rm km\ s^{-1}\ kpc^{-1}$. The details of the potential are described fully in Paper I. While using individual star particles allows for a more complex potential than a fixed analytic expression, there is a risk of numerical issues from the discreteness of the potential particles. To minimize this, we smoothed the particle gravitational contribution by adding their mass to the grid at AMR level 4, with a cell size of 50\,pc. The dark matter halo was included using a static NFW profile \citep{NFW1997}, with parameters given in Paper I. 

For the initial galaxy radial gas distribution, we assumed an exponential density profile based on the observations of M83 by \citet{Lundgren2004}. These are shown in equation (2) of Paper I, and the initial gas temperature was $10^4$ K. The gas was given a circular velocity calculated as $V_{\rm cir}(r) = ({\rm G} M_{\rm tot}(r)/r)^{1/2}$, where $M_{\rm tot}(r)$ is the enclosed mass of stars, dark matter and gas within the radius, $r$. 

%%%%%%%%%% subsection %%%%%%%%%%
\subsection{Star formation and feedback}
\label{Star formation and feedback}

Star formation and stellar feedback were included in two out of the three simulations, starting from $t = 120 \rm\ Myr$. During this first period, the global gas structure of the galaxy was forming, as gas fell into the potential created by the non-axisymmetric bar and spiral star potential particles. 120\,Myr takes the galaxy through roughly one rotation of the pattern speed and allows the gas to fully fragment.

We used a star formation and feedback algorithm based on \citet{CenOstriker1992}. A star particle forms in a grid cell when the following five criteria are met: (1) The cell's gas density exceeds $n_{\textrm{threshold}} = 1.3 \times 10^4\ \textrm{cm}^{-3}$ ($n_{\textrm{cell}} > n_{\textrm{threshold}}$). This is consistent with the density at which star formation is observed to occur \citep{Lada2010, Ginsburg2012, Padoan2014}. (2) There must be a net gas inflow into the grid cell, $\nabla \cdot {\bf v_{\rm cell}} < 0$. (3) The cooling time is less than the dynamical time ($t_{\rm cool} < t_{\rm dyn} \equiv \sqrt{3 \pi / 32 {\rm G} \rho_{\rm tot}}$) or the temperature is less than 11,000 K. (4) The star particle mass is greater than $m_{\textrm{min}} = 500\ \textrm M_{\odot}$. This last condition is primarily a numerical restraint imposed to avoid creating an excessive number of star particles that would slow the computation. However, this choice also justifies the thermal feedback from supernovae, since a stellar cluster less than 200\,$\rm M_{\odot}$ is unlikely to contain any Type II supernovae. (An assumption based on \citet{Salpeter1955} stellar initial mass function for the frequency of massive stars $> 8\ \rm M_{\odot}$.) (5) Finally, the cell must be maximally refined.

If all of these criteria are met, a star particle is created at the centre of the cell with a mass, $m_{\rm star} = m_{\mathrm{cell}} (\Delta t/t_{\mathrm{dyn}}) f_{\rm SFE}$, where $f_{\rm SFE}$ is the star formation efficiency parameter (roughly star formation per dynamical time).  The particle velocity matches that of the gas in its birth cell. The efficiency parameter was selected to be $f_{\rm SFE} = 0.002$. This value was selected to match the star formation rate of the simulation to that of M83 (see Figure~\ref{star_formation_rate} later in this paper).  This value is a factor of ten times lower than the observed GMC average star formation efficiency \citep{KrumholzTan2007} (but note that we are applying it on a cell-by-cell basis rather than averaged over an entire GMC).

If a cell does not have sufficient mass to form the particle, but otherwise fulfills the criteria for star formation, its mass is added to a global tally of unborn stellar mass. When this summation exceeds the minimum mass, a star particle is formed. This stochastic system is employed for all the stars formed in our simulation, since the low efficiency parameter prevents any one cell having sufficient mass immediately. 

While the star formation algorithm creates each star particle instantaneously, the stellar feedback takes place over an extended period of time to mimic the evolution of the cluster. The cluster is assumed to form its stars at a rate $ \propto \tau {\rm e}^{-\tau}$, where $\tau = (t - t_{\rm form})/t_{\rm dyn}$ and $t_{\rm form}$ is the formation time of the star particle. The mass of stars formed at a time $t$ with time step $\Delta t$ is therefore
\begin{equation}
\Delta m_{\rm sf} = M(t+\Delta t) - M(t)
= m_{\rm star} [(1+\tau_0){\rm e}^{-\tau_0} - (1+\tau_1){\rm e}^{-\tau_1}]
\end{equation}
where $M(t)$ is the total stellar mass formed between $t_{\rm form}$ and $t$, $\tau_0 = (t - t_{\rm form})/t_{\rm dyn}$, and $\tau_1 = (t + \Delta t - t_{\rm form})/t_{\rm dyn}$.  This newly created stellar mass is then used in the feedback routine.

Our feedback scheme adds thermal energy at each time step equivalent to $\Delta E = f_{\rm SN} (\Delta m_{\rm sf} {\rm c}^2)$, where $f_{\rm SN}$ is the fraction of the rest-mass energy of the star particle that has been converted into heat. Our value of $f_{\rm SN} = 3 \times 10^{-6}$ is equivalent of about three supernovae for every $500\ \textrm M_{\odot}$ star particle formed assuming one supernova ejects $10^{51} {\rm erg}$. The thermal energy is distributed over the 19 neighbouring cells.  To account for winds and other ejecta, $m_{\rm ej} = f_{\rm ej} \Delta m_{\rm sf}$, is subtracted from the star particle and returned as gas to the grid cell with momentum $m_{\rm ej} \bf{v_{\rm *}}$, where $\bf{v_{\rm*}}$ is the star particle velocity. The assumed fraction of mass ejected by all stars integrated over their life is taken to be $f_{\rm ej} = 0.25$.

Table \ref{run names} shows the simulations we performed. To compare the effects of the star formation and thermal stellar feedback, we performed three different runs: no star formation or feedback run (NoSF), only star formation run (SFOnly) and thermal feedback run (SNeHeat). NoSF was published in Paper I.

\begin{table}
\begin{center}
\begin{tabular}{lcc}
	Simulation & Star formation & SNe heating \\ \hline
	NoSF & No & No \\ 
	SFOnly & Yes & No \\ 
	SNeHeat & Yes & Yes \\ 
\end{tabular} 
\caption{Summary of the simulations compared in this paper. Star formation and stellar feedback are included from $t =$ 120 Myr.}
\label{run names}
\end{center}
\end{table}

%%%%%%%%%% subsection %%%%%%%%%%
\subsection{Cloud analysis}
\label{Cloud analysis}

The GMCs in our simulation were identified as coherent structures contained within contours at a threshold density of $n_{\rm gas} = 100$\,cm$^{-3}$, similar to the observed mean volume densities of typical galactic GMCs. Note that we do not include formation or destruction of molecules. Instead, we assume that the cloud would consist of both a molecular core and atomic envelope.

Our clouds were assigned to an environment group (spiral, bar or disc) based on their physical location within the galaxy. If a cloud is found within galactic radii $2.5 < r < 7.0$\,kpc, it is recognised as a spiral cloud. Bar clouds form in a box-like region at the galactic centre, with a length of 5.0\,kpc and width 1.2\,kpc that rotates with the bar potential. The nuclear region inside 600\,pc is excluded from cloud analysis due to the difficulty in accurately tracking clouds in such a high density area. Outside $r = 7.0$\,kpc, clouds are designated disc clouds. The outermost ring ($r > 8$ kpc) is formed in a Toomre instability during the fragmentation of the initial conditions and is also excluded from cloud analysis, since it is affected by our idealized initial conditions.

\begin{figure}
\centering
	\includegraphics[width=8.5cm, bb=0 0 776 720]{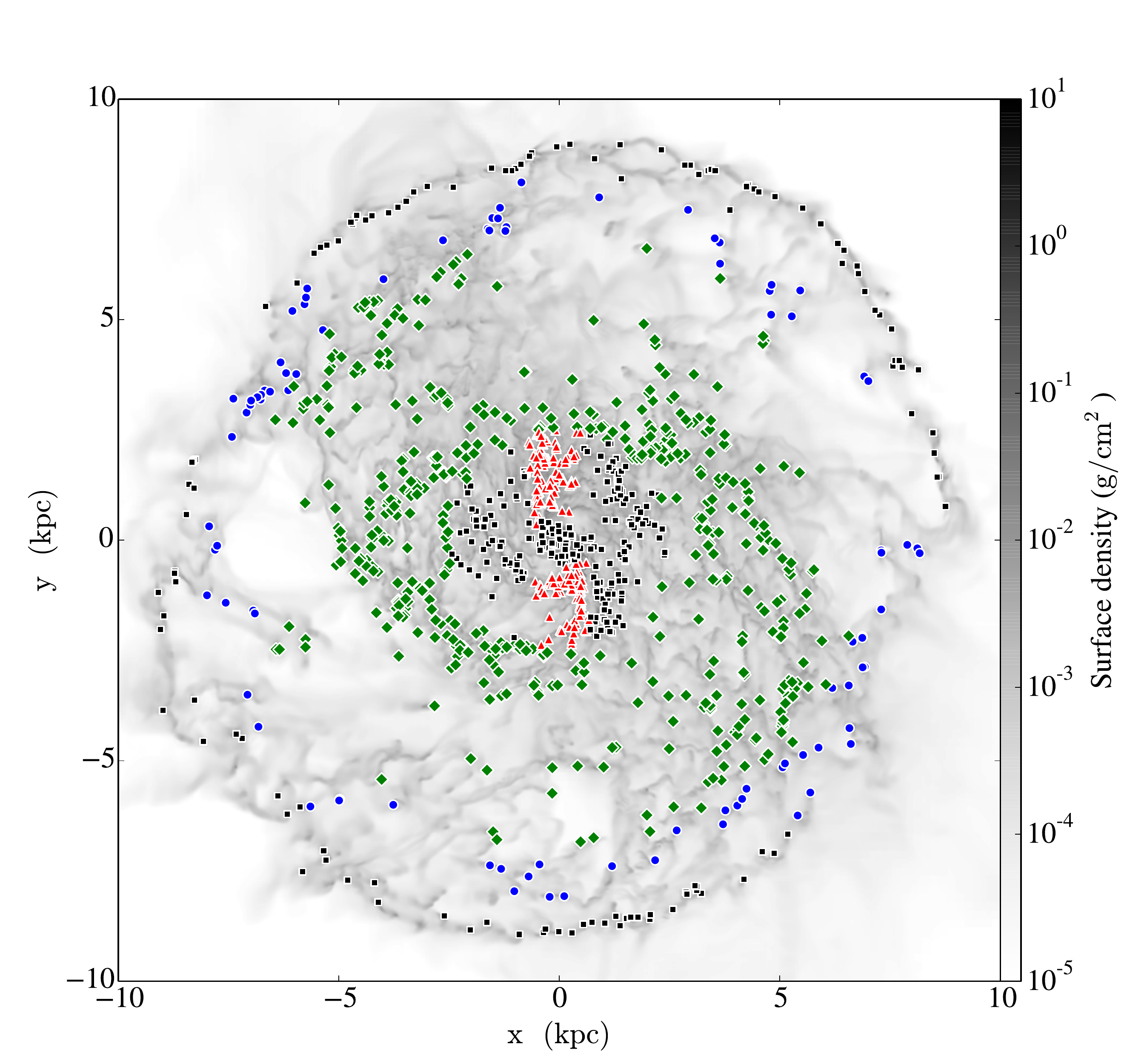}
	\caption{The location of the three different galactic environments: bar, spiral and disc. Coloured markers show the positions of the identified clouds: red triangles are clouds in the bar region, green diamonds are spiral clouds and blue circles are in the disc. The black squares show clouds not included in the analysis. The background image shows the gas surface density of SNeHeat run at $t = 200$\,Myr.}
	\label{cloud_categorisation}
\end{figure}

These three environment regions are shown in Figure~\ref{cloud_categorisation}. The monchrome background image is the surface density of the gas in run SNeHeat at $t = 200$\,Myr and is overlaid with the cloud positions, coloured to indicate their assigned environment. Blue shows clouds identified as being in the disc region, green for the spiral and red for the bar. Black clouds do not sit in any of the analysed regions and are not included in the analysis.

%%%%%%%%%% Results %%%%%%%%%%
\section{Results}
\label{Results}

%%%%%%%%%% subsection %%%%%%%%%%
\subsection{The Interstellar Medium}
\label{The stellar feedback effects on the ISM}

\begin{figure*}
\centering
	\includegraphics[width=18.0cm, bb=0 0 754 720, clip, viewport = 0 220 754 500]{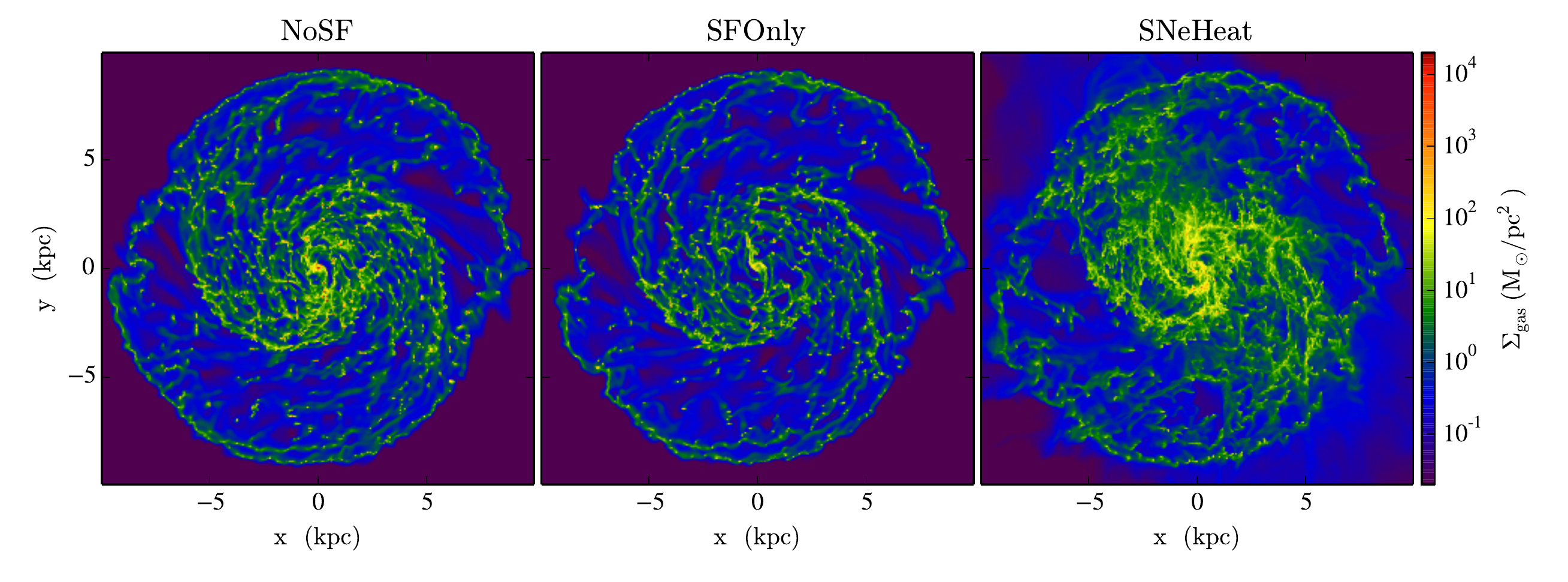}
	\caption{The global gas distribution in the galactic disc for the three runs: left is NoSF, middle is SFOnly, and right is SNeHeat. Images show the gas surface density of the face-on disc at $t =$ 200 Myr. Each image is 20 kpc across. The galactic disc rotates anticlockwise.}
	\label{density_projection_whole_galaxy}
\end{figure*}

Figure~\ref{density_projection_whole_galaxy} shows the gas distribution in the global galactic disc in all three runs at $t = 200$\,Myr. The non-axisymmetric bar and spiral pattern, rotates anticlockwise, taking 120\,Myr for one complete rotation.  

In the runs without feedback, NoSF and SFOnly, the galactic disc settles into a quasi-equilibrium state with no large structural change after about $t = 150$\,Myr. The grand design bar and spiral can be clearly seen in the left and centre panels, showing a gas distribution qualitatively similar to the lower resolution CO observations of M83 \citep{Lundgren2004}. While showing strong simularities, these two non-feedback runs are not identitical. The inclusion of star formation removes high density gas, converting it into star particles (not shown). This can be seen most clearly in the bar and spiral regions, where the gas is typically denser than in the disc. In these regions, the SFOnly run shows a clear reduction in surface density. 

The inclusion of thermal feedback in run SNeHeat, seen in the right-hand panel of Figure~\ref{density_projection_whole_galaxy}, produces a stronger global change. The galaxy disc remains structurally similar, but the green and yellow mid-density regions at $10^1 \sim 10^2\ \rm M_{\odot} pc^{-2}$ are more widely distributed in each of the bar, spiral and disc regions. There is also evidence of gas outflows, especially in the densest bar region. This is the effect of the thermal stellar feedback injecting energy into the dense cloud gas surrounding the newly formed star particles and causing it to expand. This gas is ejected from the dense clouds, increasing the density of the warm interstellar medium. In addition, comparing the structure of the bar and spiral features in SNeHeat with the other two runs, more gas appears to have been funneled towards the centre. We will return to this observation later in the cloud analysis.

\begin{figure*}
\centering
	\includegraphics[width=15.0cm, bb= 0 0 754 720]{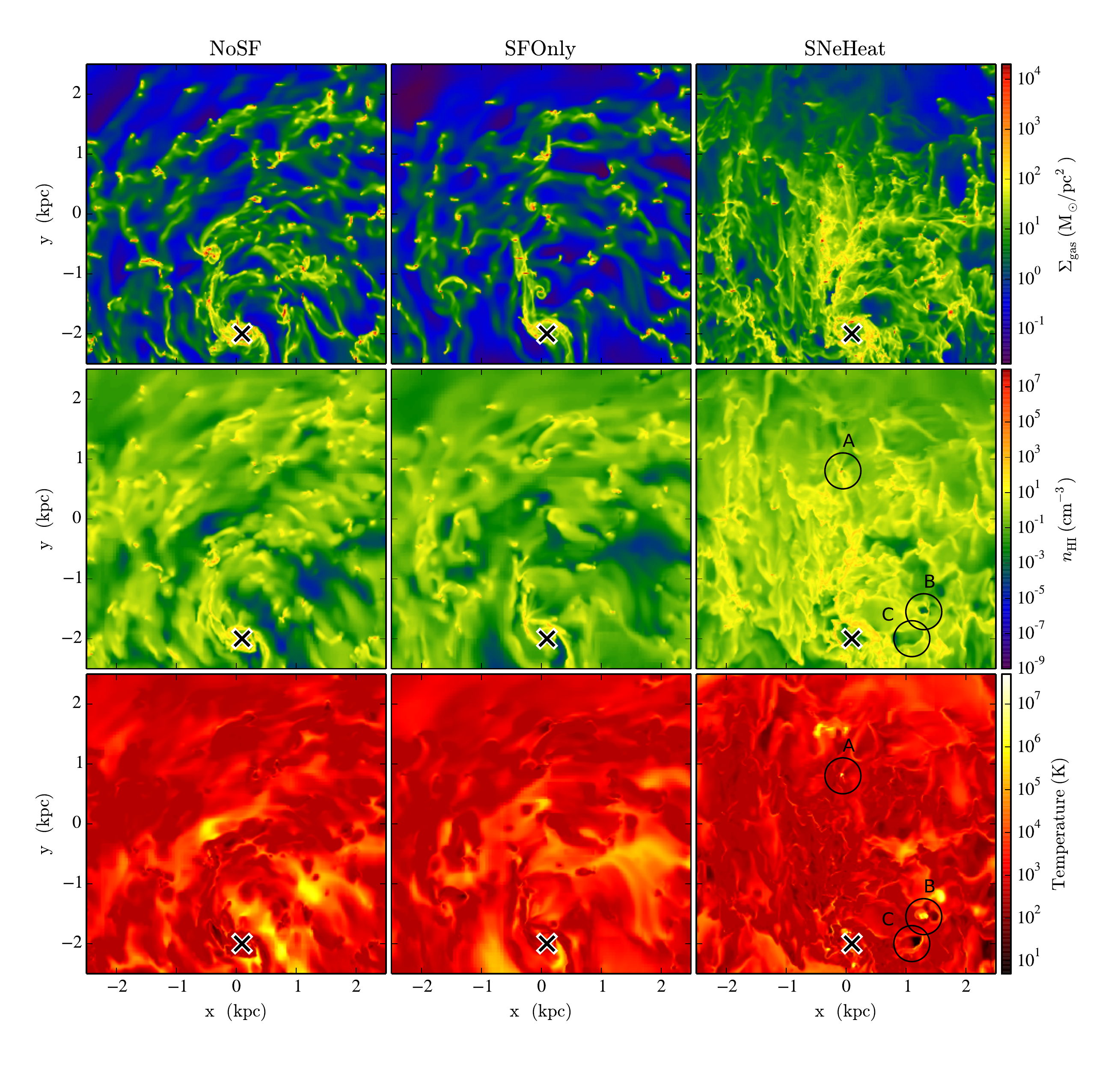}
	\caption{Close-up images of the bar-end region of the galactic disc at $t = 200$\,Myr in the three simulations. The left-hand column shows the disc for the NoSF simulation, middle is SFOnly and the right most column is for SNeHeat. Each image is a 5\,kpc across. Top to bottom, the images show the gas surface density, the gas volume density in the disc midplane and the gas temperature in the midplane. The marked `x' is the location of the galactic centre. The three circles in the right-hand panels are clouds at various stages at evolution, as discussed in the text.}
	\label{multi_plot_zoom_up_sigma_rho_T}
\end{figure*}

Figure~\ref{multi_plot_zoom_up_sigma_rho_T} shows 5\,kpc $\times$ 5\,kpc close-up images of the face-on view of the bar end at $t = 200$\,Myr, for all three simulations. The top row is the gas surface density, middle is the gas volume density in the disc mid-plane, and the bottom row is the mid-plane temperature. In the left-hand column, the NoSF run shows the gas evolution without the influence of star formation or feedback. There are clearly defined dense knots of material corresponding to the clouds. These show up as red collapsed regions in the density panels and dark red cold gas in the temperature panels. These clouds are undergoing tidal interactions in the denser environments, pulling spiral filaments of structure around them. Closest to the galaxy centre, in the bar itself, more massive clouds are forming and dominating the local gravitational environment. This increases the density of the tidal tails to produce a transient population of clouds, as discussed in more detail in Paper I. In between the densest regions of cloud material at the beginning of the spiral arms is warm, low-density gas. 

The SFOnly simulation in the middle column shows a similar structure of dense knots, but noticeably fewer tidal filaments. This is due to the star formation eroding the reservoirs of dense gas to shrink the size of the largest clouds. The result is a less extended population of massive structures to pull gas away from neighbouring clouds. 

The effect of feedback in the SNeHeat images (right-hand column of Figure~\ref{multi_plot_zoom_up_sigma_rho_T}) has made even more significant differences. The gas surface density images have lost their ordered clump and filament structure seen in NoSF and SFOnly, and instead show a higher density reservoir of gas surrounding the clouds. While the inclusion of star formation has reduced the number of massive clouds and tidal tails, adding stellar feedback has dispersed part of the gas in the clouds to form a new, more turbulent mix of filaments and clumps. In the mid-plane slice of density and temperature for run SNeHeat, three circles show the early (circle A), mid (circle B) and late (circle C) phase of a feedback site. In the centre of circle A, we see a high density knot of gas that has risen to high temperatures. This is where a star particle has just injected thermal energy into the surrounding gas. This thermal energy causes the gas to expand, leading to a hot cavity that can be seen inside circle B. The thermal energy is eventually radiatively lost, leading a high density rim to the expanding cavity, as can be seen in circle C. 

\begin{figure*}
\centering
	\hspace{-15pt}
	\subfigure{
	\includegraphics[width=12.0cm, bb=0 0 754 720, clip, viewport = 0 220 754 500]{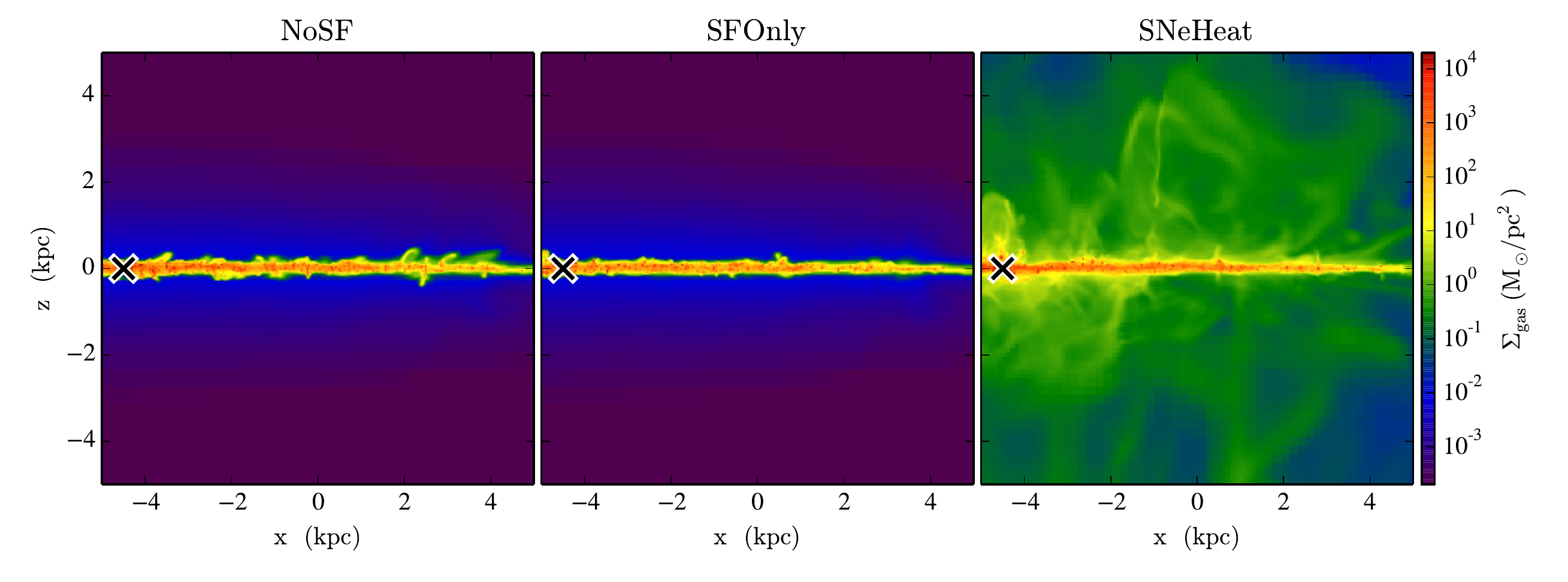}}
	\hspace{-15pt}
	\subfigure{
	\includegraphics[width=6.0cm, bb=0 0 576 432]{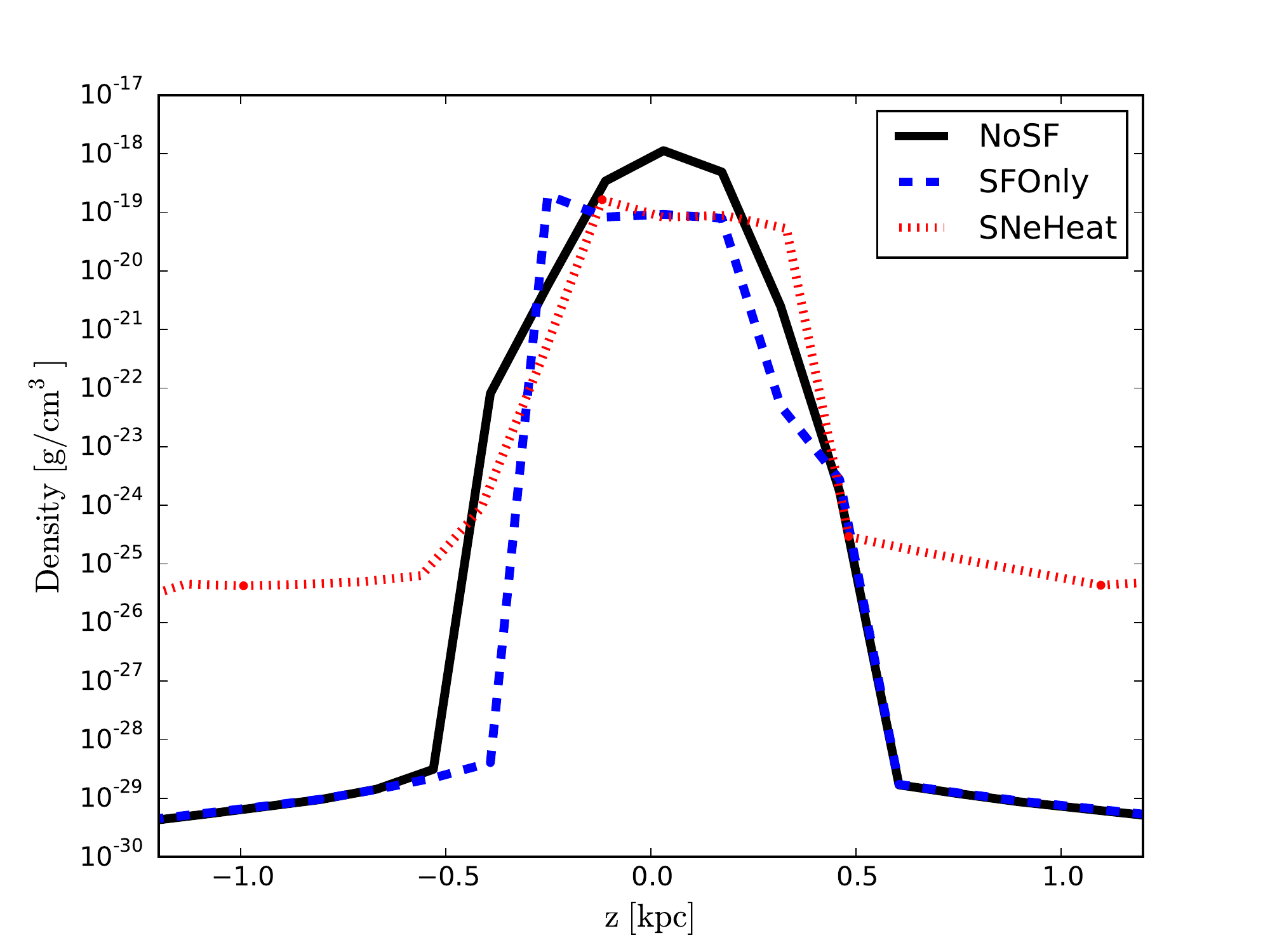}}
	\caption{Left: Edge-on gas surface density of the galactic disc at $t =$ 200 Myr for the three runs: NoSF, SFOnly, and SNeHeat. Each image is 10\,kpc across. The `x' mark at the left side of the image shows the galactic centre. Right: 1D profile of the scale height for the galactic disc at $t = 200$\,Myr. The y-axis is the mass weighted average density as a function of the position, $z$. The black solid line is NoSF, the blue dashed line is SFOnly, and the red dotted line is SNeHeat.}
	\label{edge_on_disc}
\end{figure*}

The effect of the different stellar physics on the vertical profile of the disc is shown in Figure~\ref{edge_on_disc}. The left-hand side of the figure shows the projected density along the disc edge for each of the three runs, while the right-hand side is the disc scale height. Without thermal feedback, the vertical height of the disc is primarily controlled by cloud interactions that can scatter denser material off the disc plane. This effect is more marked in the NoSF run images compared to the SFOnly for the same reasons that fewer filaments were seen in Figure~\ref{multi_plot_zoom_up_sigma_rho_T}; the star formation results in fewer high mass clouds that are the most efficient at promoting cloud interactions and tidally stripping other clouds. This difference results in a small change in the scale height shown in the right-hand plot, where the disc height for SNOnly is marginally lower than for the NoSF run. 

The very densest gas in the SNeHeat run extends to a similar height as that in NoSF and SFOnly, with a scale height of about 400\,pc. However, there is a large difference in the lower density gas above and below the disc.  While the density drops sharply beyond 500\,pc in NoSF and SFOnly, the density in the SNeHeat run is $10^4$ times higher. Visually, plumes of gas are being ejected from the disc by the thermal stellar feedback to form a galactic fountain that stretches up to several kiloparsecs above the mid-plane. The fact that the densest gas seems largely unaffected suggests that the thermal feedback is having the strongest affect on the medium density warm ISM, in keeping with the very extended filamentary structure seen in Figure~\ref{multi_plot_zoom_up_sigma_rho_T}.

Since M83 is a face-on galaxy, its scale height cannot be measured. Our value of about 400\,pc for the dense gas compares favourably to other galaxies, e.g. $100 \sim 500$\,pc in the Milky Way \citep{Lockman1984, Sanders1984}, 180\,pc in LMC \citep{Padoan2001}, 200\,pc in NGC 891 \citep{Scoville1993}.

\begin{figure*}
\centering
	\includegraphics[width=15.0cm, bb=0 0 737 640, clip, viewport = 0 220 630 470]{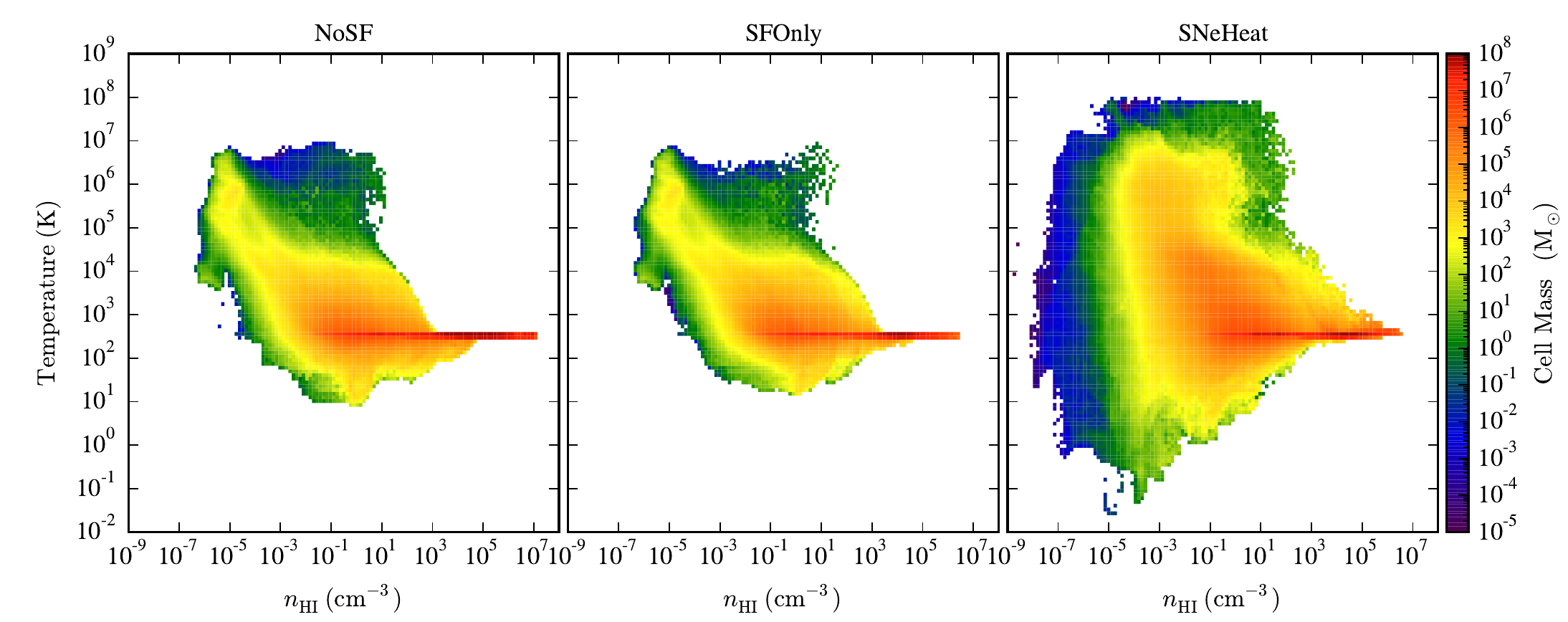}
	\caption{The ISM mass distribution in temperature versus number density at 200 Myr. Left is NoSF, middle is SFOnly, and right is SNeHeat.}
	\label{multi_phase_plots}
\end{figure*}

The structural differences in the ISM can also be seen clearly in Figure~\ref{multi_phase_plots}, which shows the two-dimensional phase diagrams of temperature versus density in the three runs. All three simulations show a continuous distribution of densities and temperatures, demonstrating that the ISM phases are not distinct bodies of gas but part of a smoothly changing system. The NoSF and SFOnly simulations show almost the same ISM phase distributions: mass collects in the cold and dense clouds, sitting at the radiative cooling limit of 300\,K. The surrounding warm ISM is in rough pressure equilibrium (but at multiple pressures) at temperatures around $10^4$\,K and $0.1$\,cm$^{-3}$ and there is a small hot phase around $10^6$\,K. Below our cooling limit at 300\,K, gas above and below the disc expands adiabatically expansion to cool to about 100\,K. This is slightly more marked in the NoSF case, as the increased number of cloud interactions due to the formation of more massive clouds scatters more gas off the mid-plane. Likewise, the conversion of gas into star particles reduces the amount of very dense gas in the SFOnly run. 

Compared to the NoSF and SFOnly ISM, the addition of thermal feedback greatly changes the distribution of the warm and hot gas. Thermal stellar feedback is injected into the densest star forming regions, producing hotter gas at cloud densities above $100$\,cm$^{-3}$. As with the disc images in Figures~\ref{multi_plot_zoom_up_sigma_rho_T} and \ref{density_projection_whole_galaxy}, the main difference is seen at lower densities outside the clouds, between $10^{-3} \sim 10^1\ \mathrm{cm^{-3}}$. The warm ISM now contains more gas mass and spreads to higher densities, reflecting the heavy tangle of filaments visible in Figure~\ref{multi_plot_zoom_up_sigma_rho_T}. At densities below $< 10^{-1}\ \mathrm{cm^{-3}}$, we see a much broader range of temperatures corresponding to the galactic fountain, which throws hot gas off the disc where it cools and returns to the midplane, as seen in the edge-on disc views of Figure~\ref{edge_on_disc}.

\begin{figure*}
\centering
	\includegraphics[width=13.0cm, bb=0 0 576 432]{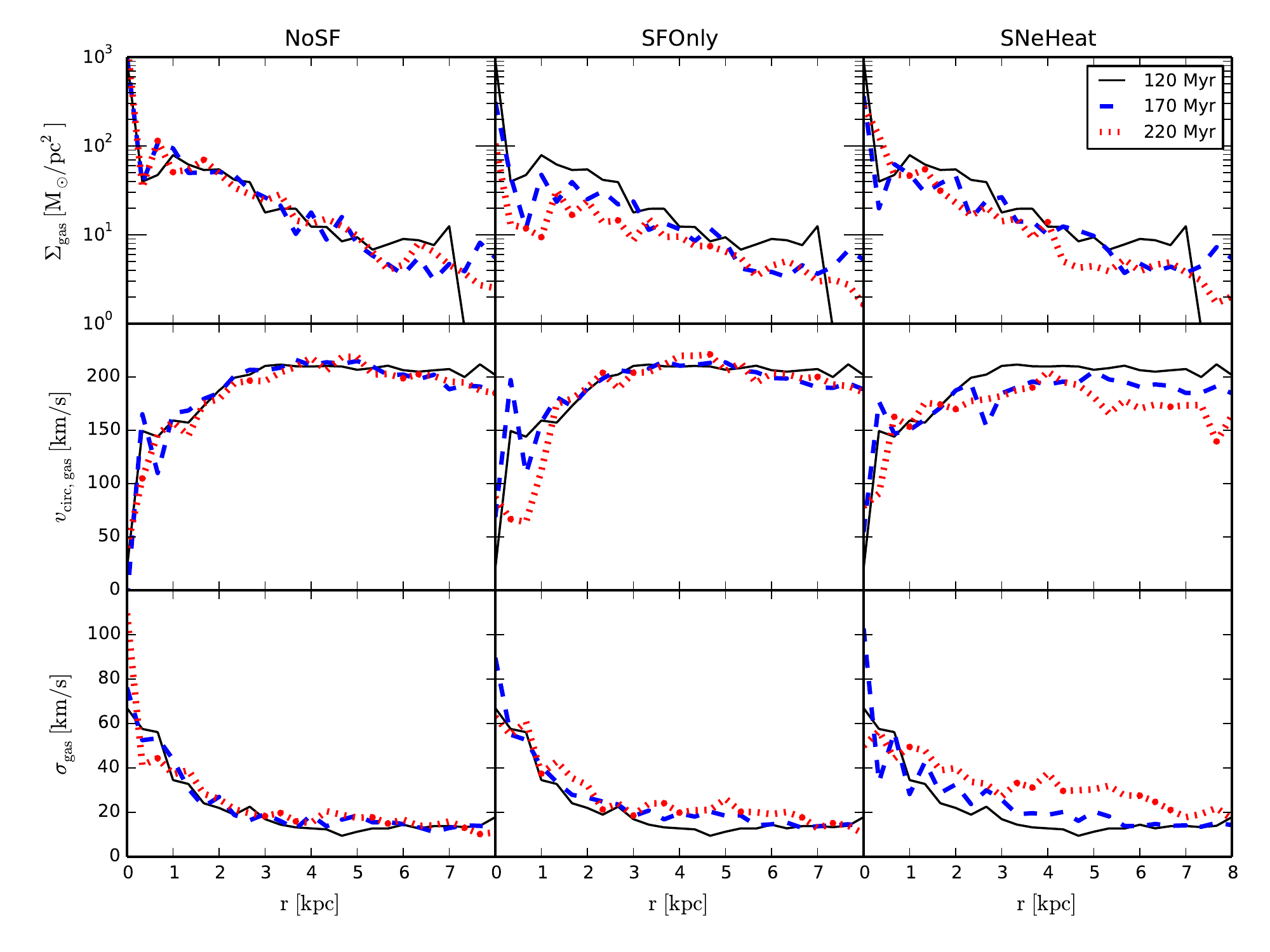}
	\caption{Radial profiles of the evolution of the three galactic discs at $t = 120, 170, 220$\,Myr. Bins are azimuthally averaged with a width of 333\,pc. Top to bottom rows show (1) the gas surface density, $\Sigma_{\rm gas} = \int_{-1\mathrm{kpc}}^{+1\mathrm{kpc}}\rho(z)\mathrm{d}z$, (2) gas circular velocity (mass-weighted average over $-1\ \mathrm{kpc} < z < 1\ \mathrm{kpc}$) and (3) 1D gas velocity dispersion defined by $\sigma_{\rm gas} = \sqrt{(\mbox{\boldmath $v$} - \mbox{\boldmath $v$}_{\rm cir})^2/3}$ (also mass-weighted average over $-1\ \mathrm{kpc} < z < 1\ \mathrm{kpc}$). Star formation and feedback is included only {\it after} $t = 120 \rm\ Myr$, so that the profiles of the three runs at 120 Myr are the same in each column.}
	\label{multiplot_radial_profile}
\end{figure*}

The evolution of the ISM is shown in the one dimensional radial profiles in Figure~\ref{multiplot_radial_profile}. At 120\,Myr, all galaxy discs have the same profile, since this is just prior to the beginning of star formation and feedback, as described in Section~\ref{Star formation and feedback}. For the NoSF run, all three profiles in gas surface density, circular velocity and velocity dispersion remain steady over time. Once star formation in include in SFOnly, the gas surface density drops with time as gas is removed from the disc to form stars. This is most pronounced in the galactic bar region ($r < 3\ \mathrm{kpc}$) where the star formation timescale is shortest, due to the high gas surface density. Once feedback is included in SNeHeat, the same decrease in gas surface density is seen in the outer parts of the disc beyond $\sim 4$\,kpc, but the the central region of the galaxy remains gas rich. This suggests one of two possibilities are occurring. The first option is that star formation is being suppressed more strongly in the central region by the thermal feedback than in the outer parts of the disc. The second possibility is that gas is inflowing to the galactic centre, replenishing the material lost due to star formation. This latter option is supported by the images of the global galaxy disc in Figure~\ref{density_projection_whole_galaxy}, where the gas surface density appears to be raised in the central region compared to the SFOnly and NoSF runs. Our exploration of the star formation in the next section will confirm this view. 

The middle and bottom rows show the changes in circular velocity and velocity dispersion over time. The velocity dispersion rises in the central $< 2$\,kpc of all discs due to the elongated elliptical motion of gas in the bar potential. The densely packed bar region is also the site of the highest number of cloud collisions, further raising the velocity dispersion in this region. The inclusion of star formation in SFOnly produces a slight rise in the velocity dispersion as the coldest gas is removed from the disc to form stars, leaving the average to be weighted by the faster moving warm sector. This effect increases far more significantly when thermal feedback is injected in SNeHeat, heating the dense gas and creating a more extensive warm and hot ISM phase. The average circular velocity also shows changes in the SNeHeat run, implying that the predominantly circular motion of the gas is being disrupted. This again points to an increase in radial gas motions, as suggested by the possible inflow of material into the galactic center just discussed.

%%%%%%%%%% subsection %%%%%%%%%%
\subsection{Star formation}
\label{The stellar feedback effects on star formation}

\begin{figure*}
\centering
	\hspace{-60pt}
	\subfigure{
	\includegraphics[width=8.0cm, bb=0 0 576 432]{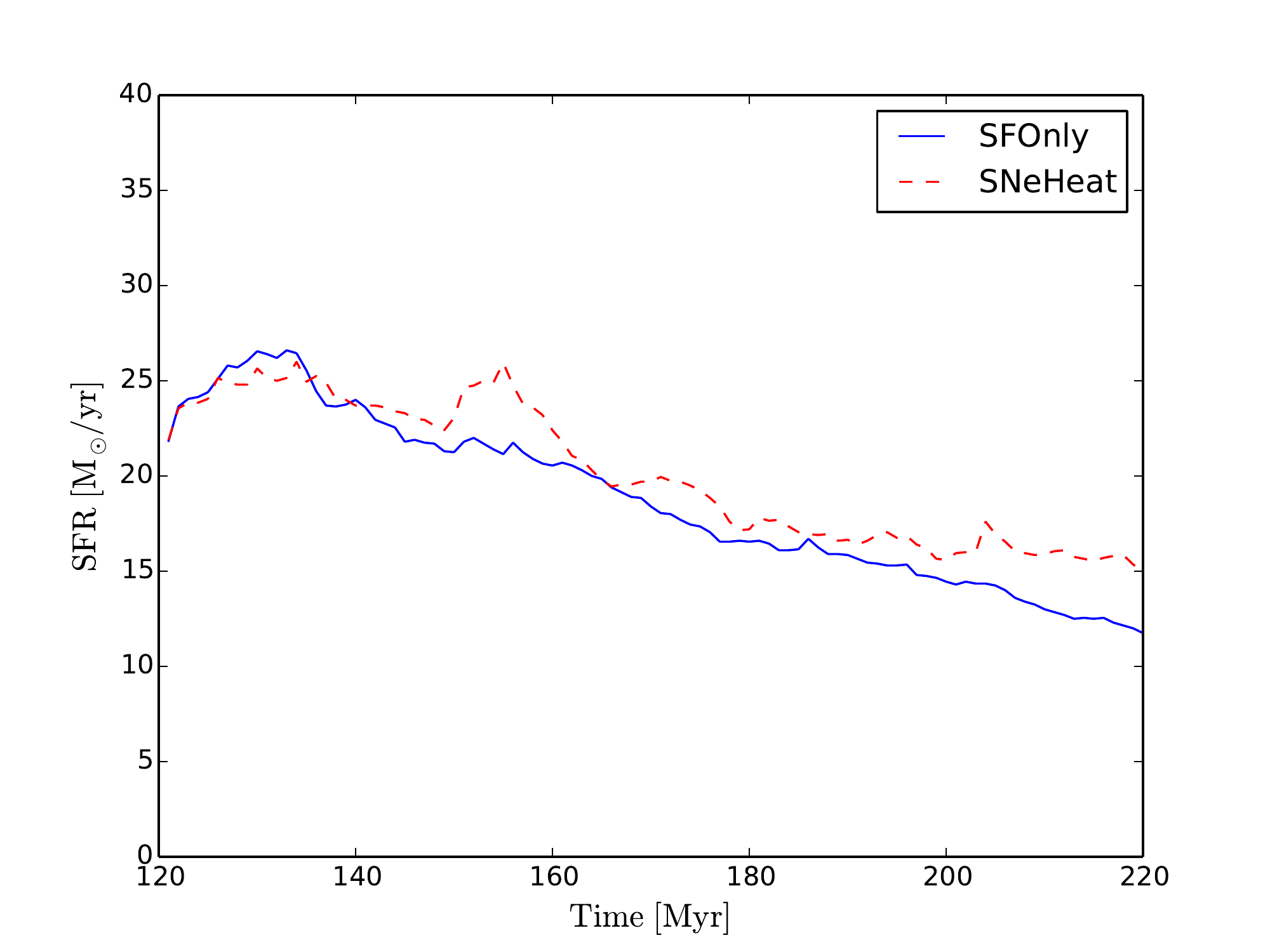}}
	\hspace{-25pt}
	\subfigure{
	\includegraphics[width=12.0cm, bb=0 0 864 432]{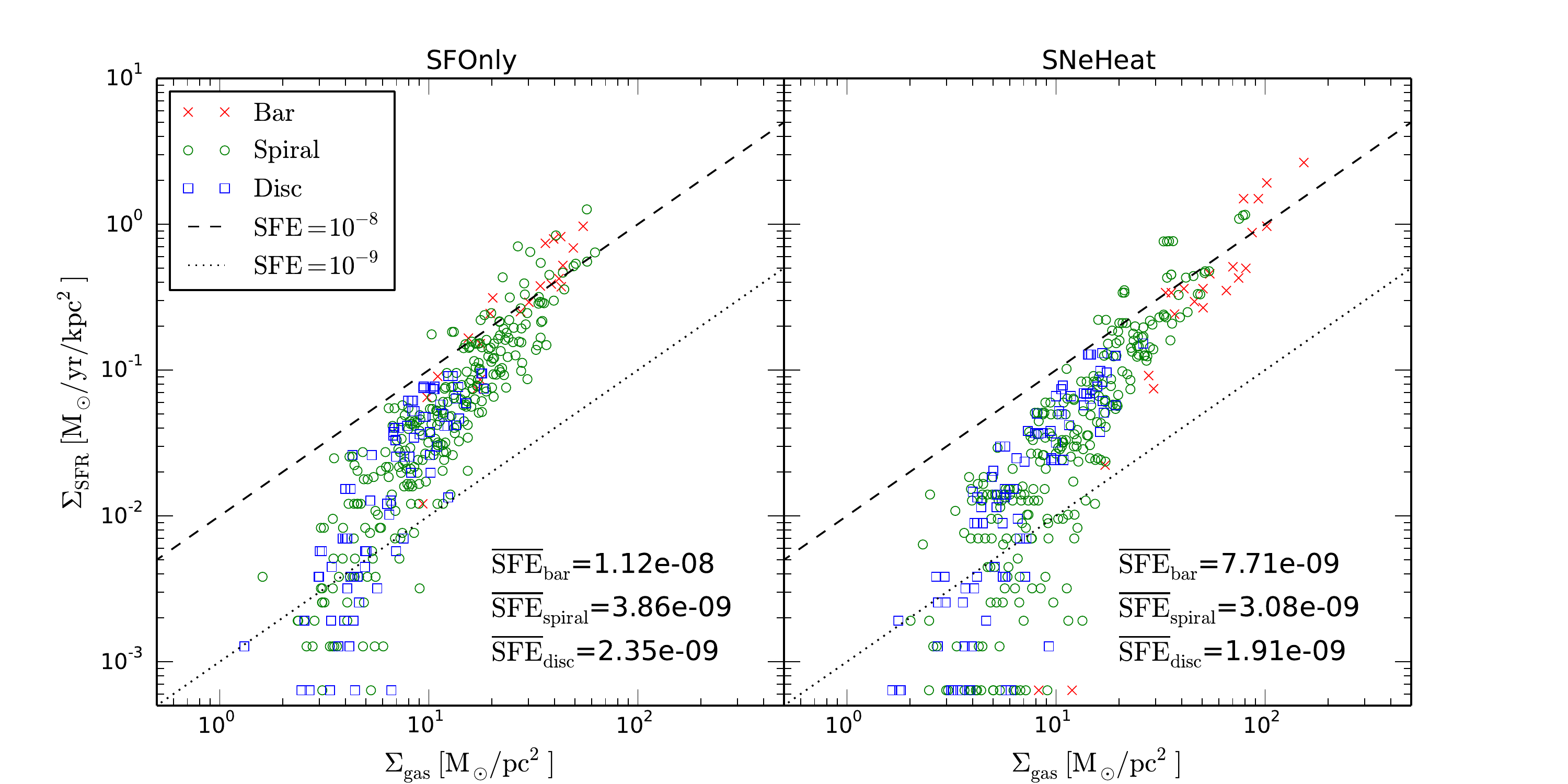}}
	\hspace{-80pt}
	\caption{The star formation rate (SFR) in our runs including active star formation, SFOnly and SNeHeat. Left shows the star formation history, beginning from when star formation is included at 120\,Myr. The right plots show the Kennicutt-Schmidt relation for SFOnly and SNeHeat at $t = 200$\,Myr.  Different coloured symbols mark regions in the bar, spiral and disc, as shown in the legend. The surface area is calculated in the face-on plane and each point corresponds to a cylindrical region of radius 500\,pc and height 5\,kpc. The black lines show constant star formation efficiency (SFE): $10^{-8}$ and $10^{-9}$ ($\mathrm{yr^{-1}}$).}
	\label{star_formation_rate}
\end{figure*}

The left panel of Figure~\ref{star_formation_rate} shows the star formation history of the galaxy in SFOnly (blue solid line) and SNeHeat (red dashed line) from when star formation begins at t=120 Myr. In the first 10\,Myr, both discs show a rising star formation rate, reaching over 25\,M$_\odot$yr$^{-1}$. This burst phase is a consequence of starting star formation in the pre-fragmented gas at $t = 120$\,Myr. The large number of dense clumps all rapidly form stars. After this starburst, the star formation rate gradually decreases as the quantity of dense gas decreases due to conversion into stars. Both the SFOnly and SNeHeat simulations show similar histories over the next 100\,Myr, decreasing to around 15\,M$_\odot$yr$^{-1}$ by 200\,Myr. This value roughly corresponds to the observed SFR of $5 \sim 20\ \mathrm{M_{\odot}yr^{-1}}$ in M83 \citep{Hirota2014}. It should be noted that this match is partially due to our choice of star formation efficiency, as described in Section~\ref{Star formation and feedback}.

Despite the similar trend, the star formation rate when feedback is included in SNeHeat is slightly higher than for SFOnly for all times after $\sim 140$\,yr. This suggests that the addition of thermal feedback may initially disrupt the star forming dense gas (between $t = 120 - 140$\,Myr) but this effect is swiftly overwhelmed by a positive feedback loop. The result is a nearly constant SFR in the last 20\,Myr of the SNeHeat simulation, while SFOnly continues to use up the available dense gas. 

This positive effect on the star formation can stem from multiple possible effects. Star formation can be triggered in shells of compressed gas in the wake of strong outflows following the thermal energy injection. Alternatively, gas may be more effectively recycled by the galactic fountain and other outflows, making the net efficiency of gas into stars higher. The ejection of gas into the warm ISM may also encourage the replenishment of gas in the galactic centre, where short dynamical times in the dense region encourage higher star formation rates. 

While triggering and recycling may play a role, the feeding of gas into the galaxy centre appears to be a strongly controlling factor. The maintenance of a high central gas density is seen in Figures~\ref{multiplot_radial_profile} and \ref{multi_plot_zoom_up_sigma_rho_T}, and the star formation history in Figure~\ref{star_formation_rate} confirms this is not due to a low star formation rate, but from a replenishment of fresh material. Feedback therefore has minimal effect on star-forming cloud gas, but boosts star formation by maneuvering gas to maintain dense areas.  

The right panel of Figure~\ref{star_formation_rate} shows the relation between the SFR surface density ($\Sigma_{\rm SFR}$) and the gas surface density ($\Sigma_{\rm gas}$), averaged in cylindrical regions of radius 500\,pc and height 5\,kpc. The colours of each region marker indicate the location in the galaxy: bar (red crosses), spiral (green circles) and disc (blue squares). The resulting correlation is known as the Kennicutt-Schmidt relation: $\Sigma_{\rm SFR} \propto {\Sigma_{\rm gas}^N}$, where $N$ is typically observed to be between 1 and 2 for combined molecular and atomic hydrogen \citep{Kennicutt1998}. Recent sub-kpc resolution observations also show an index around 1 for the pure molecular gas and slightly steeper for the atomic component \citep{Bigiel2008, Leroy2008}. Both the SFOnly and SNeHeat relations follow an index of $\sim 1.7$, in reasonable agreement with observations. 

This super-linear relation results in a trend in the star formation efficiency (SFE) in each region: bar $>$ spiral $>$ disc. Such ordering is present in both SFOnly and SNeHeat, although the thermal feedback does slightly reduce the SFE in all regions. Moreover, there are bar regions in the SNeHeat simulation with markedly higher gas surface density ($> 10^2\ \mathrm{M_{\odot}pc^{-2}}$) and SFR ($> 10^0\ \mathrm{M_{\odot}yr^{-1}kpc^{-2}}$), further adding to the picture that this region is fed by gas inflow. 

This high SFE in the bar is contrary to observations that show a lower efficiency in this region than in the spiral arms \citep{Momose2010, Hirota2014}. In \citet{Fujimoto2014b}, we showed that such a difference could be reproduced by assuming that star formation was triggered during cloud collisions using an adaption of the star formation model of \citet{Tan2000} that allowed for a dependence on cloud collision velocity as suggested by \citet{Takahira2014}. In this situation, cloud interactions in the bar region were too violent to result in productive star formation, due to the elliptical global gas motion in the stellar bar potential. This result is not replicated in this model as our star formation model is based on local gas density, rather than cloud interactions. This is presently an unfortunate necessity, since even at our resolution, we do not have the spatial accuracy to resolve a shockfront during a cloud interaction that would lead to star formation, nor is it possible to calculate the cloud interaction rate swiftly enough to be used during the hydrodynamical calculation. Our  results are therefore consistent with the estimates in Paper I, which also primarily considers gas density, but some variation in the bar region is possible if an alternative cloud interaction-based scheme were to be employed.

%%%%%%%%%% subsection %%%%%%%%%%
\subsection{Cloud properties}
\label{The stellar feedback effects on cloud properties}

\begin{figure*}
\centering
	\includegraphics[width=18.0cm, bb=0 0 864 648]{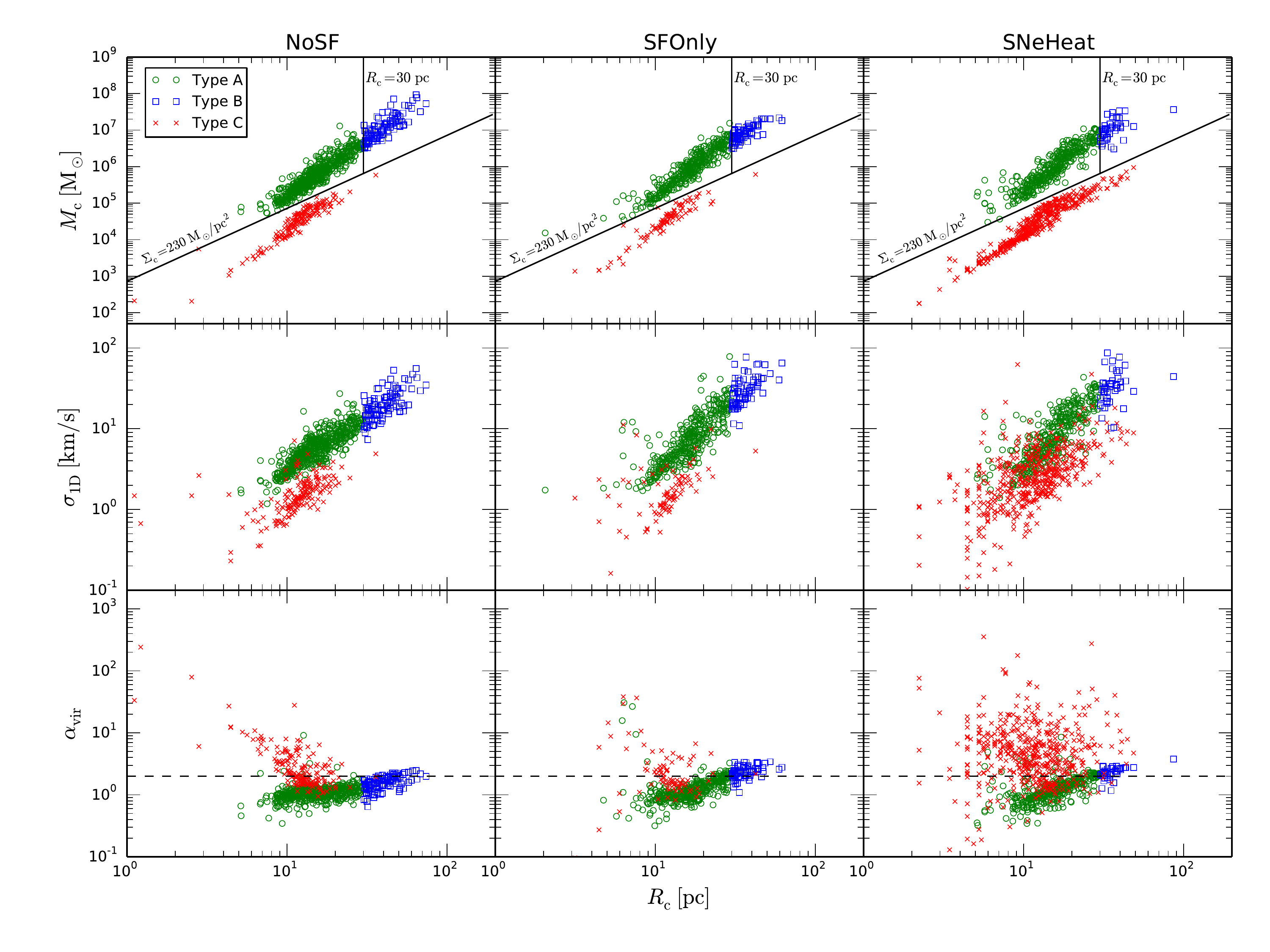}
	\caption{Scaling relations for the three types of cloud formed in the simulation. Top row is cloud mass, $M_{\rm c}$, versus the average cloud radius, computed as described in the text. This plot defines the three cloud types, with {\it Type A} clouds sitting on the upper trend of the bimodal split with surface densities $ > 230\ \mathrm{M_{\odot} pc^{-2}}$. {\it Type B} clouds lie along the same sequence, but with radii greater than 30\,pc. {\it Type C} clouds follow the lower trend with surface densities $< 230\ \mathrm{M_{\odot} pc^{-2}}$. The middle row plots the 1D velocity dispersion against cloud radius; $\sigma_{\rm 1D} = \sqrt{[(v_x-v_{{\rm c},x})^2+(v_y-v_{{\rm c},y})^2+(v_z-v_{{\rm c},z})^2]/3}$, where $(v_{x}, v_{y}, v_{z})$ is the velocity of the gas and $(v_{{\rm c},x}, v_{{\rm c},y}, v_{{\rm c},z})$ is the cloud's centre of mass velocity. The bottom row shows the viral parameter versus cloud radius; $\alpha_{\rm vir} = 5(\sigma_{\rm 1D}^2 + {c_{\rm s}}^2)R_{\rm c}/\{\mathrm{G}(M_{\rm c}+M_{\rm s})\}$, where $c_{\rm s}$ is the sound speed. The virial parameter is a measure of gravitational binding; a value greater than 2 indicates that the cloud is gravitationally unbound.}
	\label{scaling_relations}
\end{figure*}

Clouds in our simulations were defined as coherent structures of gas contained within contours at a threshold density of $n_{\rm gas} = 100$\,cm$^{-3}$ as described in Section~\ref{Cloud analysis}. In this section, we discuss the effects of star formation and stellar feedback on these cloud properties.

In Paper I, we introduced three different types of clouds that form in the simulation. The most common {\it Type A} clouds constitute the largest fraction of clouds in each galactic environment. Their properties are typical of those observed in observations; a peak mass of $5\times10^5 M_{\odot}$, radius 15 pc, and velocity dispersion of 6 km/s. The larger {\it Type B} clouds are giant molecular associations (GMAs) that formed through successful mergers between smaller clouds and had radii greater than 30\,pc. The third, {\it Type C}, clouds are transient clouds that are gravitationally unbound with short lifetimes, typically below 1\,Myr. They form in the dense filaments and tidal tails caused by cloud interactions, most commonly between the massive {\it Type B} clouds and {\it Type A}. Within the NoSF disc, Paper I found that the fractions of each cloud type depended on the frequency of interactions between clouds. This made their relative quantities environment dependent, with the bar having the highest fraction of the merger-induced {\it Type B} and the resultant {\it Type C} clouds. 

\subsubsection{Cloud scaling relations}

Figure~\ref{scaling_relations} shows three relationships between a cloud's radius and its mass (top row), one dimensional velocity dispersion (middle row) and virial parameter measuring gravitational binding (bottom row).  The cloud mass, $M_{\rm c}$, is a sum of the mass in each cell within the cloud, while the cloud radius is defined as $R_{\rm c} = \sqrt{(A_{xy} + A_{yz} + A_{zx})/3\pi}$, where $A_{xy}$ is the projected area of the cloud in the $x$-$y$ plane, $A_{yz}$ is that in the $y$-$z$ plane, and $A_{zx}$ is in the $z$-$x$ plane. The top row also shows the definitions for the three kinds of clouds.  The clouds lie on two sequences of surface density above and below 230\,M$_\odot$pc$^{-3}$. The bottom trend is the transient {\it Type C} clouds, while the top trend consists of the typical {\it Type A} clouds and the giant associations, {\it Type B}, with sizes larger than 30 pc. These definitions are the same as in Paper I. All three runs show the same split into three cloud types, ensuring these are good tracers of environmental change in the presence of feedback.

The distributions for the common {\it Type A} clouds do not change significantly with the addition of star formation, with the radius extending between $5 \sim 30$\,pc for all runs and peak values sitting at 13, 17 and 13\,pc for runs NoSF, SFOnly and SNeHeat, respectively. The mass also shows little variation. The mass range runs between $10^4 \sim 10^6 M_{\odot}$ in all three simulations and the peak cloud mass is $5 \times 10^5 M_{\odot}$ for NoSF, increasing by a factor of two in the SFOnly and SNeHeat runs to give a mass of $1 \times 10^6 M_{\odot}$. This small rise towards higher values is because a few of the giant {\it Type B} clouds that sit close to the 30\,pc divide become large {\it Type A} clouds as star formation reduces their gas content, boosting the larger {\it Type A} end of the distribution.  These values are slightly larger than the GMCs observed in the Milky Way and M33, which have peak masses of $5\times 10^4$\,M$_\odot$ and $1\times 10^5$\,M$_\odot$ respectively \citep{Roman-Duval2010, Rosolowsky2003}. However, since we do not consider molecular gas separately from atomic gas in the simulations, the observed values should be doubled to include the atomic envelope \citep{Blitz1990, Fukui2009} and may even be higher by an order of 2 or 3 to allow for uncertainties in areas such as survey resolution \citep{Benincasa2013, Hughes2013}. The mass range for our clouds from $10^4 \sim 10^7$\,M$_\odot$, comfortably includes the observed peak for these two galaxies. At present, there is no observational survey of M83, but this may change with ALMA.  

The reduction in the size of the {\it Type B} clouds can be seen in the maximum size they reach in the three simulations. Without star formation, the maximum cloud size is $10^8 M_{\odot}$ and 80\,pc. Once star formation converts these gas-rich associations into stars, the maximum size becomes $2 \times 10^7 M_{\odot}$ and 60\,pc in SFOnly. The {\it Type B} clouds are most sensitive to this effect, as their high mass results in a high star formation rate, as will be later shown. Once feedback is included, there is a similar stunting of the {\it Type B} clouds, which reach a maximum mass of  $3 \times 10^7 M_{\odot}$ and 50\,pc, with the exception of one outlier, which lies in a particularly crowded region in the global design. Compared to the SFOnly run, the {\it Type B} cloud masses show more scatter, indicative of outflows that may remove part of their diffuse outer envelope to leave a more compact object with a similar mass.  The total number of {\it Type B} clouds is also the smallest here, suggesting that, at least in some cases, enough mass is lost to convert some of the {\it Type B} clouds into {\it Type A}.

\begin{figure*}
\centering
	\includegraphics[width=8.0cm, bb=0 0 720 432]{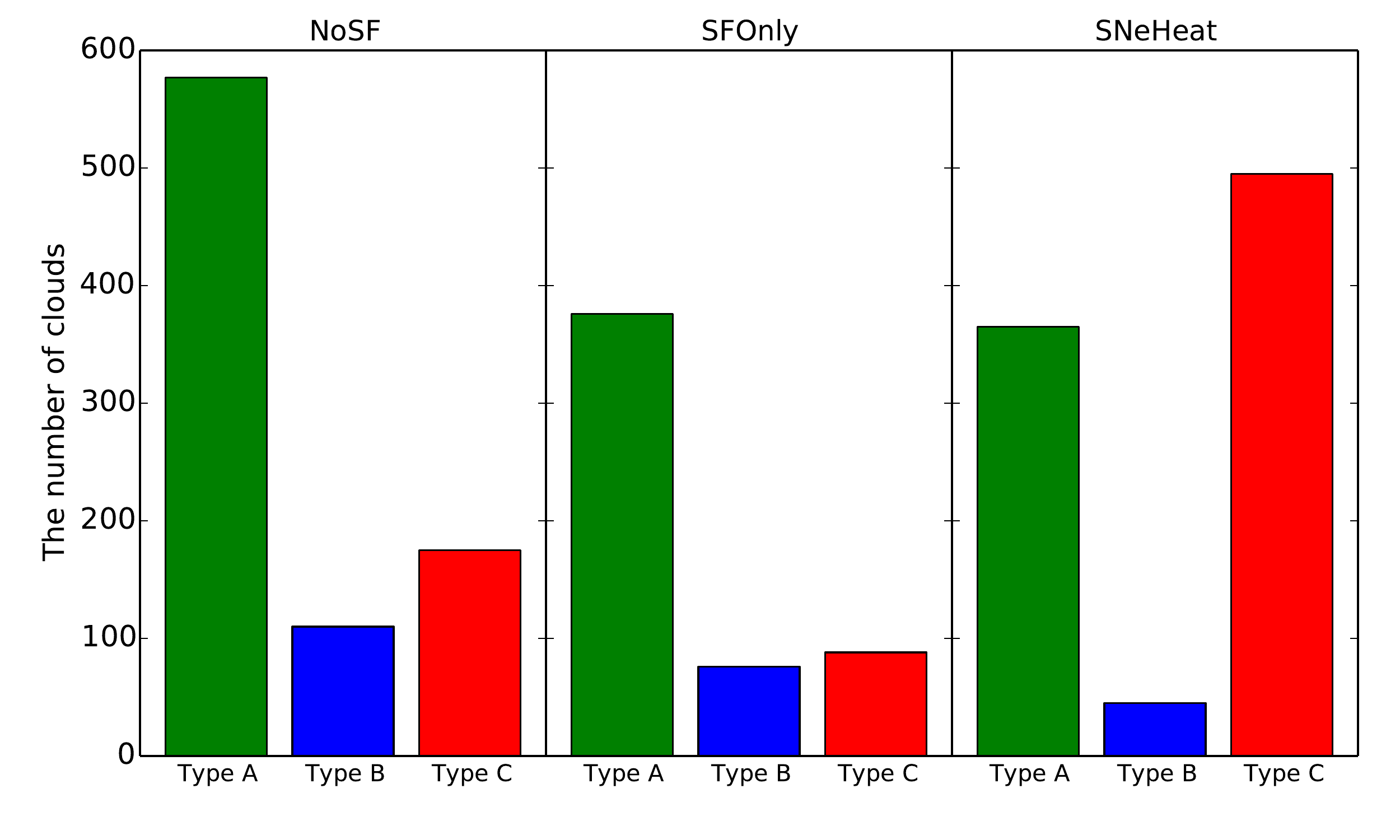}
	\caption{The number of each cloud type for all three simulations at $t =$ 200 Myr.}
	\label{numbers_cloudtype_between_3runs}
\end{figure*}

The most prominent difference across the three runs occurs for the transient {\it Type C} clouds. The cloud mass-radius distribution shows a similar trend in all three runs, but the number of {\it Type C} clouds drops markedly when star formation (but not feedback) is included. This is due primarily to the formation mechanism for this cloud type. As the number and mass of the giant GMA {\it Type B} clouds reduces with the addition of star formation, the amount of gas in tidal tails also declines, reducing the environment where {\it Type C} transient clouds predominantly form. 

This situation reverses once feedback is introduced. In the cloud-mass relation of the SNeHeat run, the number of {\it Type C} clouds blooms, forming a more extended trend that has both lower mass and higher mass tails. This can be seen most clearly in the bar chart in Figure~\ref{numbers_cloudtype_between_3runs}. The red bar shows the number of {\it Type C} clouds in each run, with the number dropping to roughly half its value between NoSF and SFOnly, but increasing by a factor of 2.5 from NoSF to SNeHeat, becoming the most dominant cloud type, by number, in that simulation. The presence of a large number of {\it Type C} clouds without a strong {\it Type B} cloud population suggests an alternative formation mechanism in the SNeHeat simulation that does not involve cloud production within tidal tails. The addition of the thermal stellar feedback boosts not only the warm ISM as discussed in Section~\ref{The stellar feedback effects on the ISM}, but also produces a population of {\it Type C} clouds in this intercloud material. Since these {\it Type C} clouds are formed in a different environment to those in the tidal tails of NoSF and SFOnly, their properties are not completely identical, with a larger number having smaller and larger radii, as seen in the scaling relations in Figure~\ref{scaling_relations}.

Figure~\ref{numbers_cloudtype_between_3runs} also confirms the slight reduction in {\it Type B} clouds seen in the cloud-mass relation for SNeHeat. These clouds have lost their outer layers to become {\it Type A} clouds, while the smaller end of the {\it Type A} clouds have been dispersed. This keeps the total number of {\it Type A} clouds roughly constant between runs SFOnly and SNeHeat. 

The middle and bottom panels of Figure~\ref{scaling_relations} show the scaling relations for the 1D velocity dispersion and the virial parameter versus radius. The mass-weighted 1D velocity dispersion of a cloud is defined as: 
\begin{align}
\sigma_{\rm 1D} = \sqrt{\frac{(v_x-v_{{\rm c},x})^2+(v_y-v_{{\rm c},y})^2+(v_z-v_{{\rm c},z})^2}{3}},
\end{align}
where $(v_{x}, v_{y}, v_{z})$ is the velocity of the gas, and $(v_{{\rm c},x}, v_{{\rm c},y}, v_{{\rm c},z})$ is the velocity of the cloud's centre of mass. The virial parameter is defined as
\begin{align}
\alpha_{\mathrm{vir}} = 5\frac{(\sigma_{\rm 1D}^2 + {c_{\rm s}}^2)R_{\rm c}}{\mathrm{G}(M_{\rm c} + M_{\rm s})},
\end{align}
where $c_{\mathrm{s}}$ is the sound speed, $M_{\rm c}$ is the total gas mass of the cloud, and $M_{\rm s}$ is total mass of star particles that are in the cloud's boundary. The virial parameter, $\alpha_{\mathrm{vir}}$, is a measure of gravitational binding, with a value less than 2 indicating that the cloud is gravitationally bound \citep{BertoldiMcKee1992}. 

The main features of the  $\sigma_{\mathrm{1D}}$-$R_{\rm c}$ and $\alpha_{\mathrm{vir}}$-$R_{\rm c}$ relations are reproduced in all three runs for all cloud types: The {\it Type A} and {\it Type B} clouds lie on the same trend with increasing velocity dispersion and virial parameter value with radius. The massive GMA {\it Type B} clouds are borderline gravitationally unbound, in keeping with their higher velocity dispersion, while the smaller {\it Type A} clouds tend to sit just on the other side of the cut-off, and are borderline bound. The transient {\it Type Cs} follow a similar trend with velocity dispersion and radius, but with a lower velocity dispersion than the larger {\it Type A} and {\it B} clouds. They lie in a different part of the $\alpha_{\mathrm{vir}}$-$R_{\rm c}$ relation, being generally small and unbound, but becoming borderline bound as their masses increases.

While the overall properties are similar, the addition of star formation and feedback does have some impact. The gradient of the  $\sigma_{\mathrm{1D}}$-$R_{\rm c}$  correlation steepens when star formation is included in SFOnly and SNeHeat. This is due to the densest parts of the cloud now forming stars, leaving the remaining gas with a higher average velocity dispersion. Without stellar physics, the peak value for the velocity dispersion is $\sigma_{\mathrm{1D}} = 6$\,km/s  but rises to 10\,km/s in SFOnly and SNeHeat. These numbers are all close to what is observed for GMCs, with M33 clouds having a characteristic velocity dispersion of 6\,km/s and the Milky Way, a lower value of 1\,km/s. 

This increase in slope for the velocity dispersion also affects the virial parameter, steepening the $\alpha_{\mathrm{vir}}$-$R_{\rm c}$ relation in SFOnly and SNeHeat compared to NoSF as well. The clouds remain borderline gravitationally bound, in keeping with the observations of clouds in the Milky Way.

\begin{figure*}
\centering
	\includegraphics[width=12.0cm, bb=0 0 720 432]{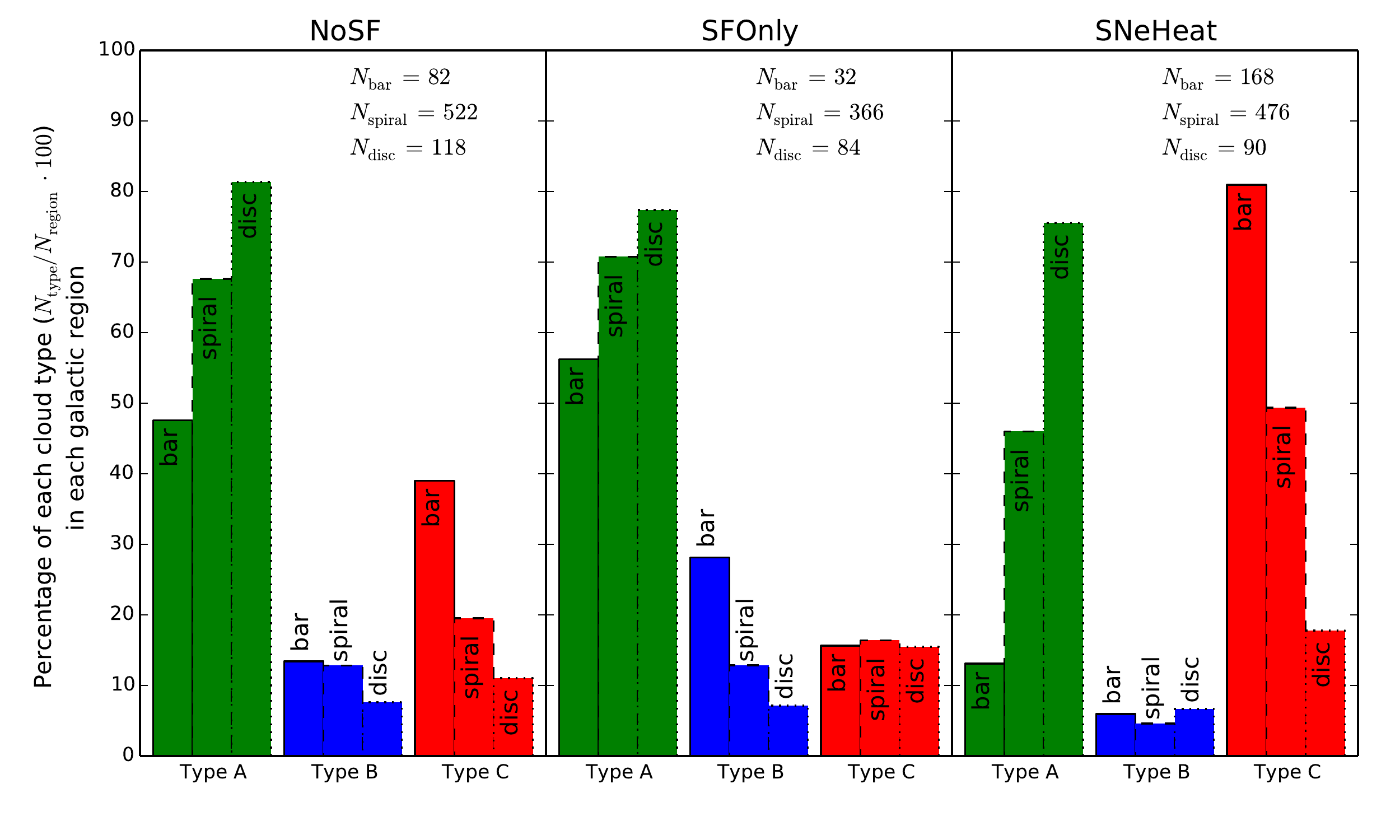}
	\caption{The percentages of each cloud type in each galactic region ($= N_{\mathrm{type}}/N_{\mathrm{region}}\times 100$) at $t =200$\,Myr.}
	\label{percentages_cloudtype_between_3regions}
\end{figure*}

As for the $M_{\rm c}$-$R_{\rm c}$ relation, the distribution of {\it Type C} clouds shows the largest difference between the simulations. Between NoSF and SFOnly, the trend for these transient clouds remains similar, with less clouds present when star formation has reduced the size of the {\it Type B}s and thus reduced the filament environment. With the addition of feedback, the number of {\it Type C} clouds is substantially larger and the scatter in their relation also increases.  The additional formation process in the denser warm ISM dispersed by stellar feedback produces a wider range of radii and velocity dispersions, leading to a more variable (yet still mostly unbound) virial parameter. 

\subsubsection{Cloud lifetime and merger rate}

The cloud lifetime and merger rate were discussed extensively in Paper I, so we include only a brief discussion here.
The distributions of lifetimes for each cloud type are very similar between the three runs: all clouds have a typical lifetime of less than 10 Myr, which agrees well with estimates that suggest lifespans of 1-2 dynamical times, with ages in the range 5-30 Myr \citep{Blitz2007, Kawamura2009, Miura2012}. The maximum {\it Type A} cloud lifetime ranges up to 30 Myr; by contrast, {\it Type B} clouds are longer lived, with a few per cent living longer than 40 Myr. The smallest lifetimes are for {\it Type C} clouds, the vast majority of which live only a few Myr.

As discussed in Paper I, the cloud lifetime is linked to interactions between clouds. The massive {\it Type B} clouds have the highest merger rate, as high as 1 merger every 2-3 Myr. They undergo many mergers during their long lifetime, accounting for their large mass and size. On the other hand, the transient {\it Type C} clouds have the lowest merger rate, in keeping with their short lifetimes. They either merge or their high virial parameter causes them to dissipate shortly after birth.

The lack of difference in the cloud lifetime and merger rate between the three runs shows that stellar feedback does not disperse the clouds (entirely), instead cloud interaction primarily controls their lifespan.

\subsubsection{Cloud Properties By Environment}

The difference in cloud properties in the three galactic environments can be seen clearly in Figure~\ref{percentages_cloudtype_between_3regions}. This bar chart shows the percentage of each cloud type in the bar, spiral and disc for the three simulations at 200\,Myr, showing how the three types are divided. In both NoSF and SFOnly, the galaxy is dominated throughout by the {\it Type A} clouds. The other two types are most prevalent in the bar region, where the high cloud interaction rate due to the elliptical gas motions encourages the formation of GMAs and the tidal tails that give rise to {\it Type C} clouds. This is followed by the spiral region, where gas is gathered in the spiral-arm potential also boosts cloud interactions, while the quiescent disc region shows the smallest percentage of {\it Type B} and {\it Type C} clouds. (Paper I contains a more detailed discussion of these differences.) Run SFOnly does show a smaller fraction of {\it Type C} clouds in the bar region than NoSF, due to the smaller {\it Type B} population.

With thermal feedback included, the percentage of {\it Type A} clouds remains high, but in the central bar and spiral regions of the disc, it is overtaken by the fraction of {\it Type C} clouds. Now forming in the dispersed gas from feedback, {\it Type C} clouds make up a larger percentage of the cloud population in every environment, but become markedly more prominent towards the galaxy centre. In contrast to this, the percentage of massive {\it Type B} clouds becomes lower and more uniform across the three environments, reinforcing our claim that they are not the primary cause of the boost in the {\it Type C's}. These large GMAs are kept smaller than in other runs by star formation and stripping of their outer layers through feedback. This stripped gas is funnelled towards the galaxy bar to keep the star formation highest in the central regions. This feedback results in an increase in the amount of feedback in the bar and spiral environments, boosting the dense filamentary structure of the warm ISM and forming more {\it Type C} clouds. 

\begin{figure*}
\begin{center}
\begin{tabular}{ccc}
	\subfigure{
	\includegraphics[width=60mm, bb=0 0 396 252]{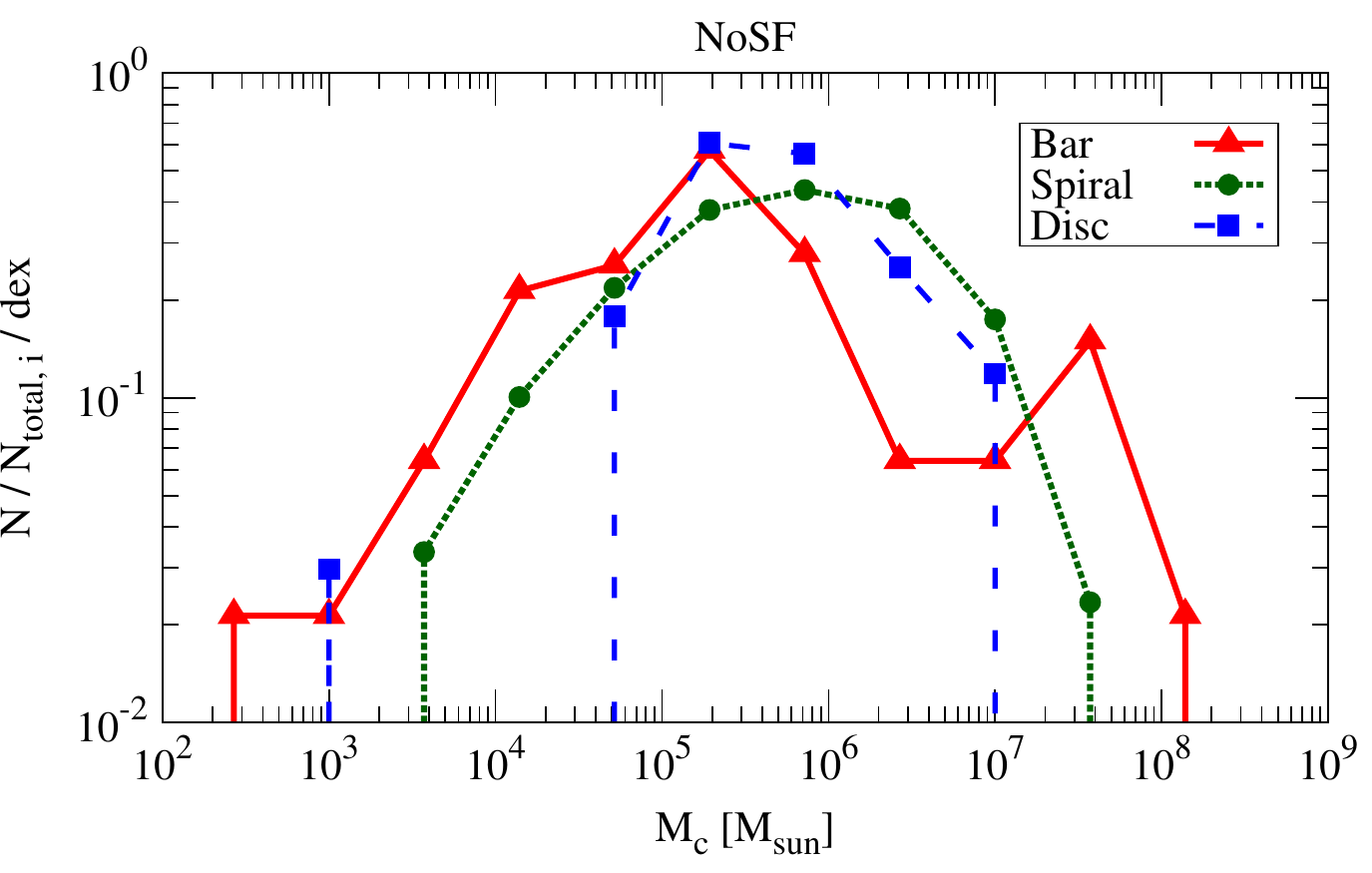}}
	\subfigure{
	\includegraphics[width=60mm, bb=0 0 396 252]{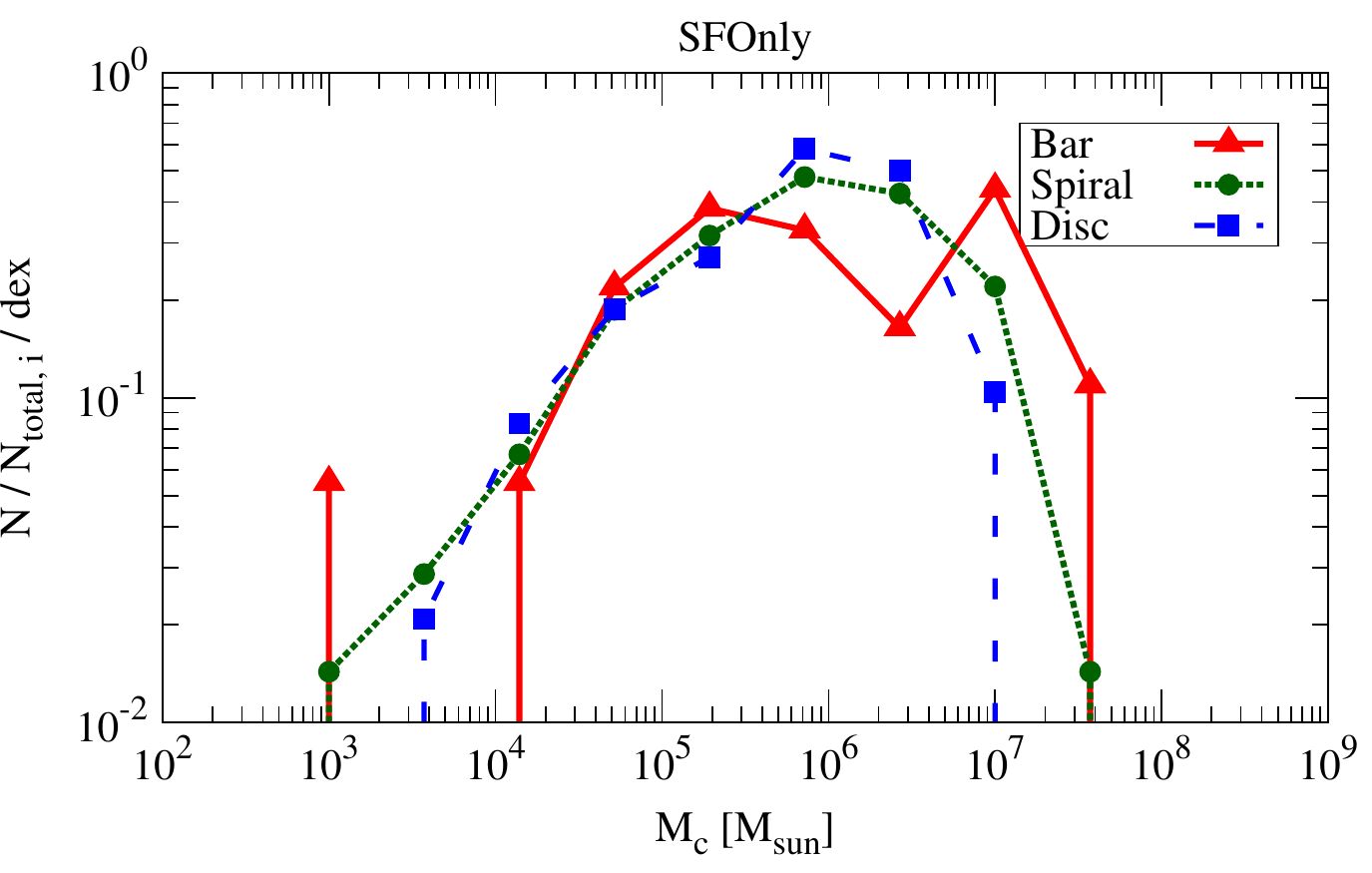}}
	\subfigure{
	\includegraphics[width=60mm, bb=0 0 396 252]{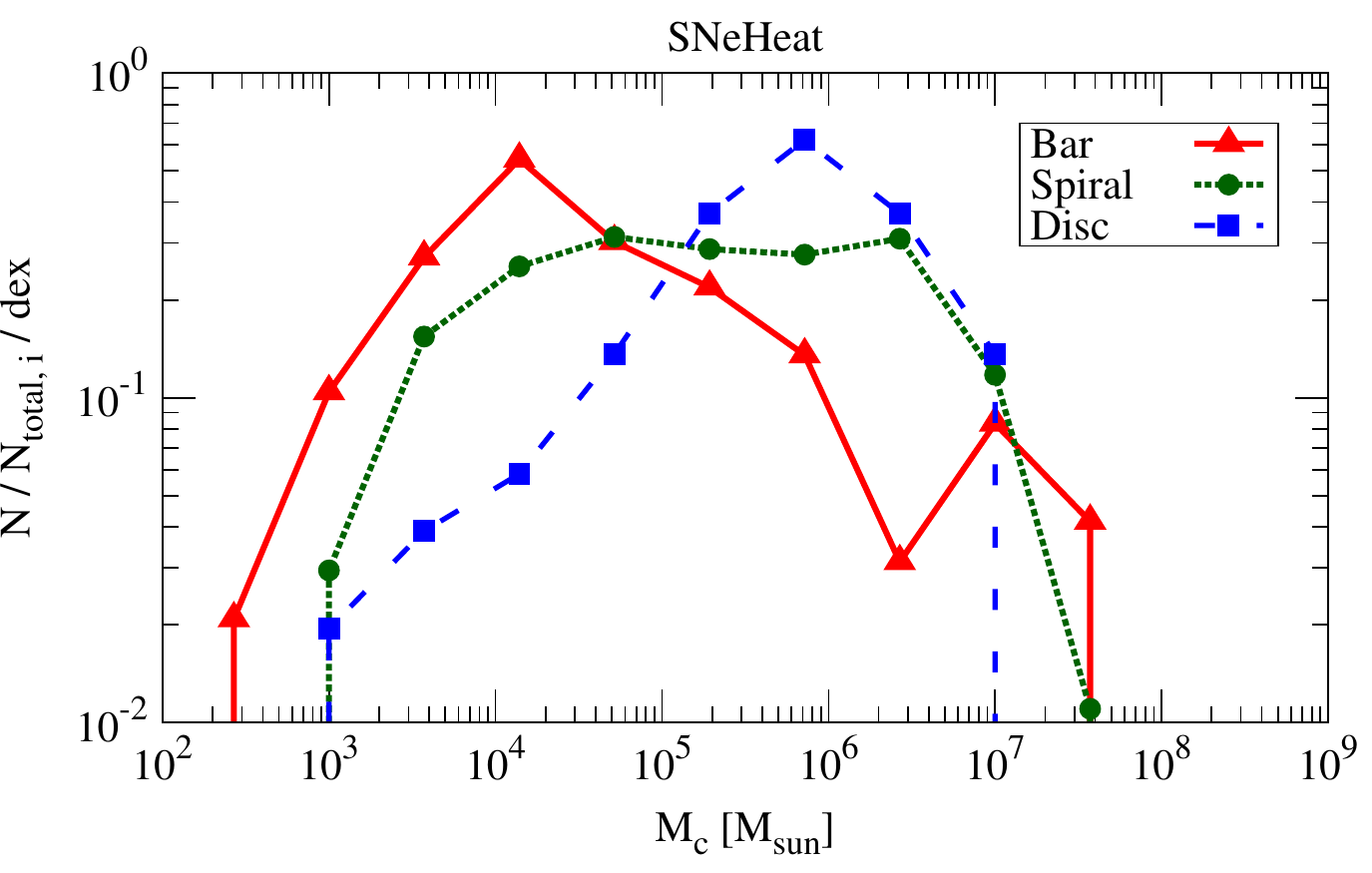}}
\end{tabular}
\begin{tabular}{ccc}
	\subfigure{
	\includegraphics[width=60mm, bb=0 0 396 252]{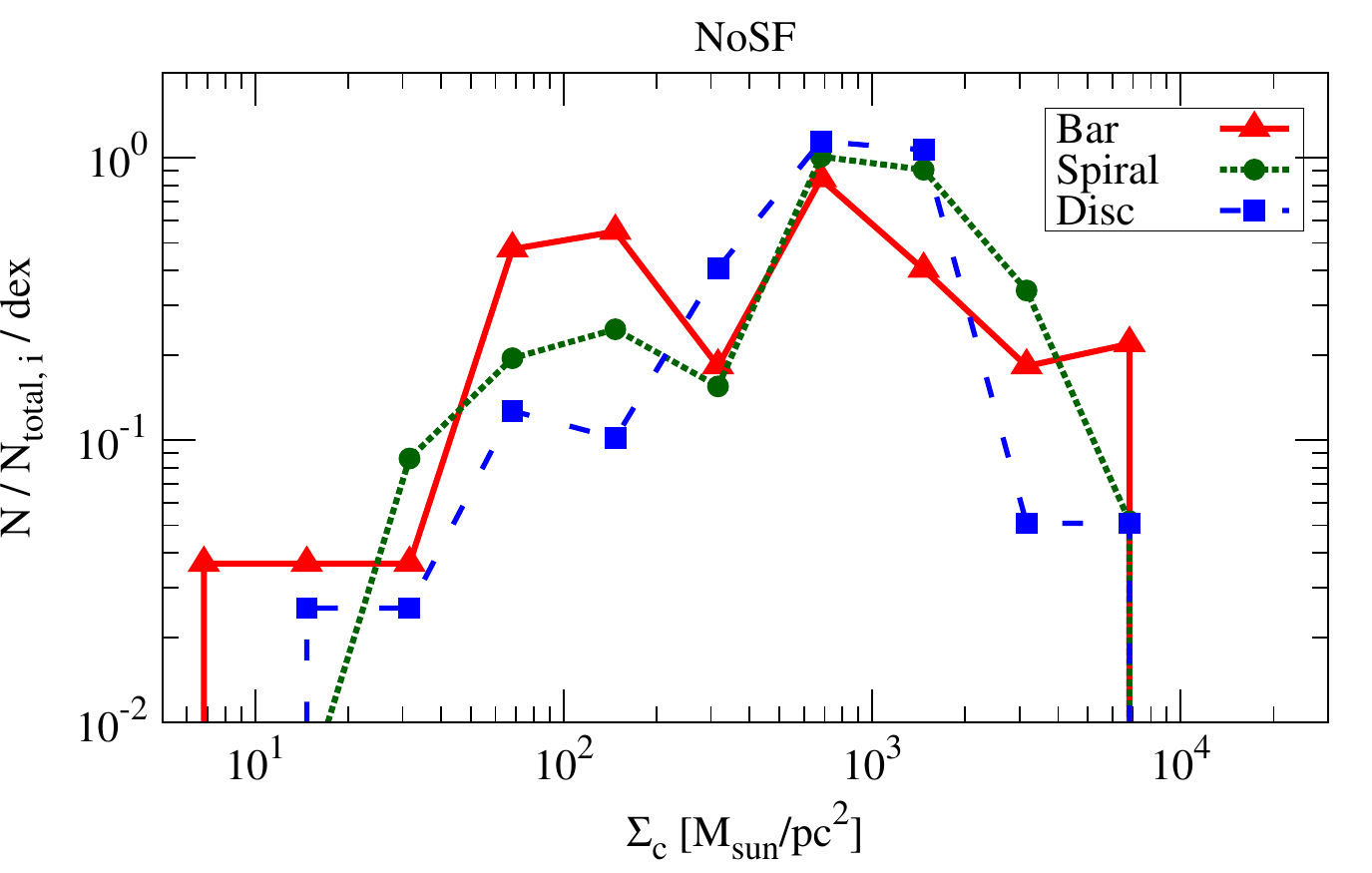}}
	\subfigure{
	\includegraphics[width=60mm, bb=0 0 396 252]{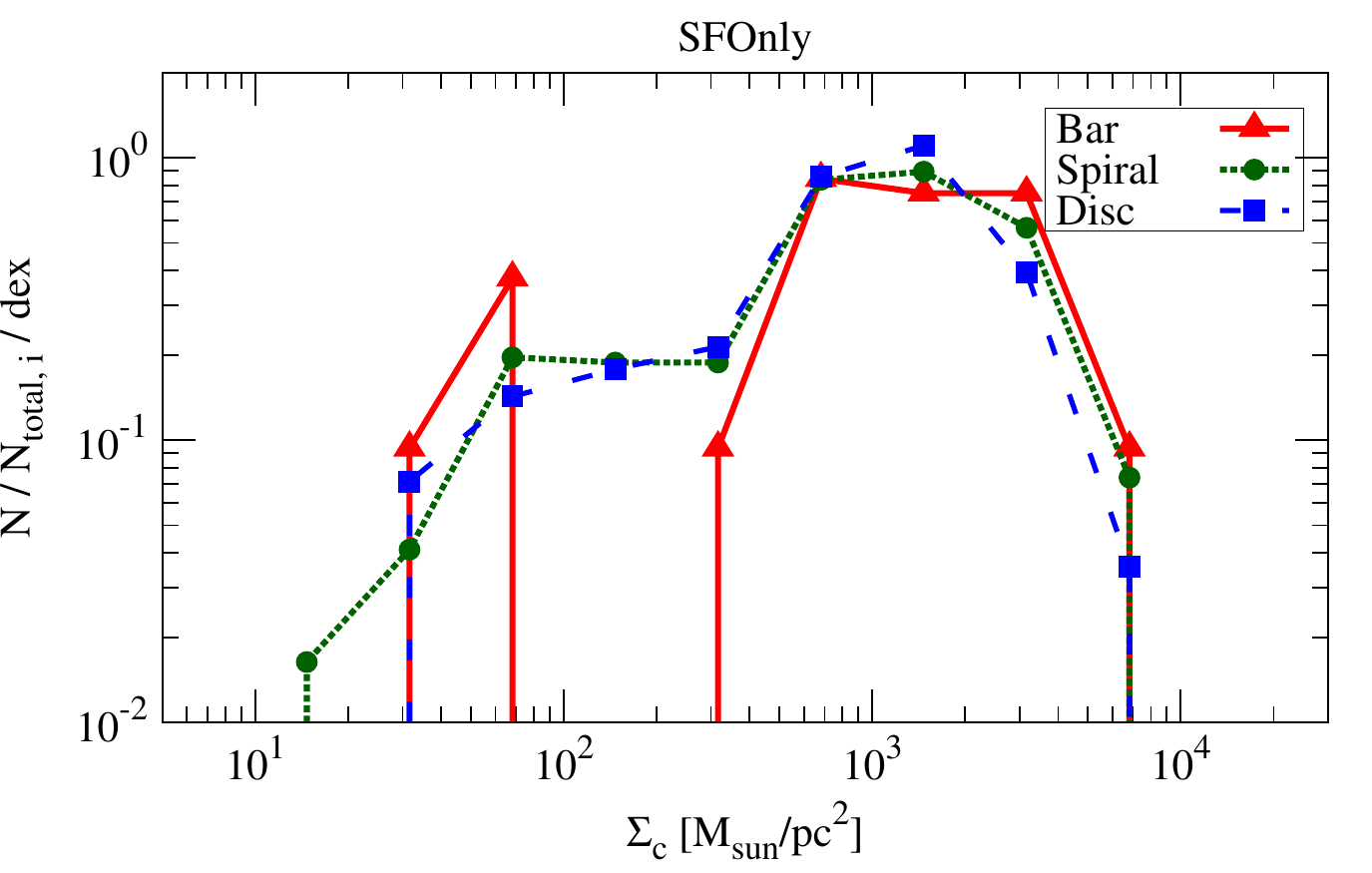}}
	\subfigure{
	\includegraphics[width=60mm, bb=0 0 396 252]{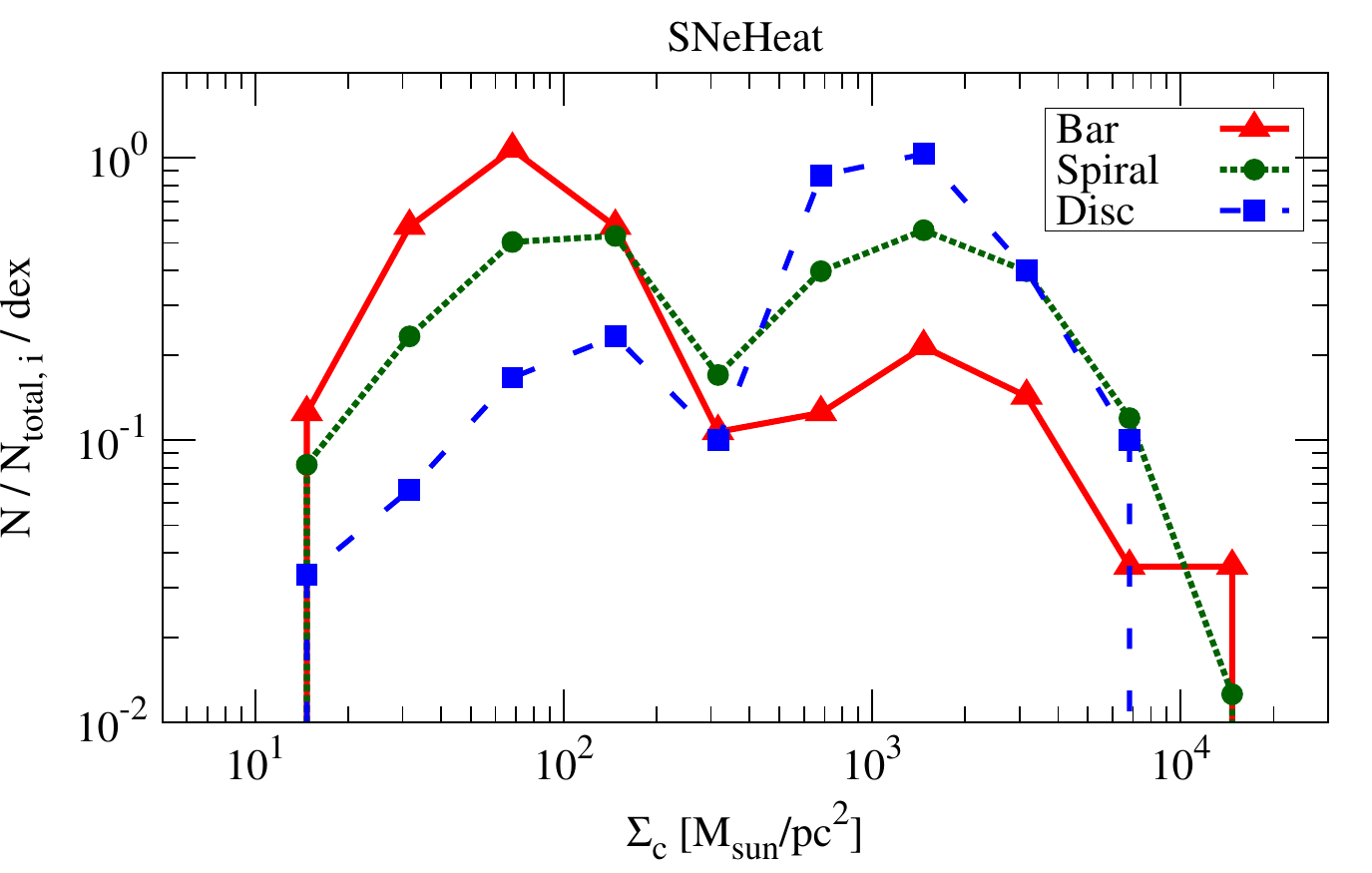}}
	\end{tabular}
\begin{tabular}{ccc}
	\subfigure{
	\includegraphics[width=60mm, bb=0 0 396 252]{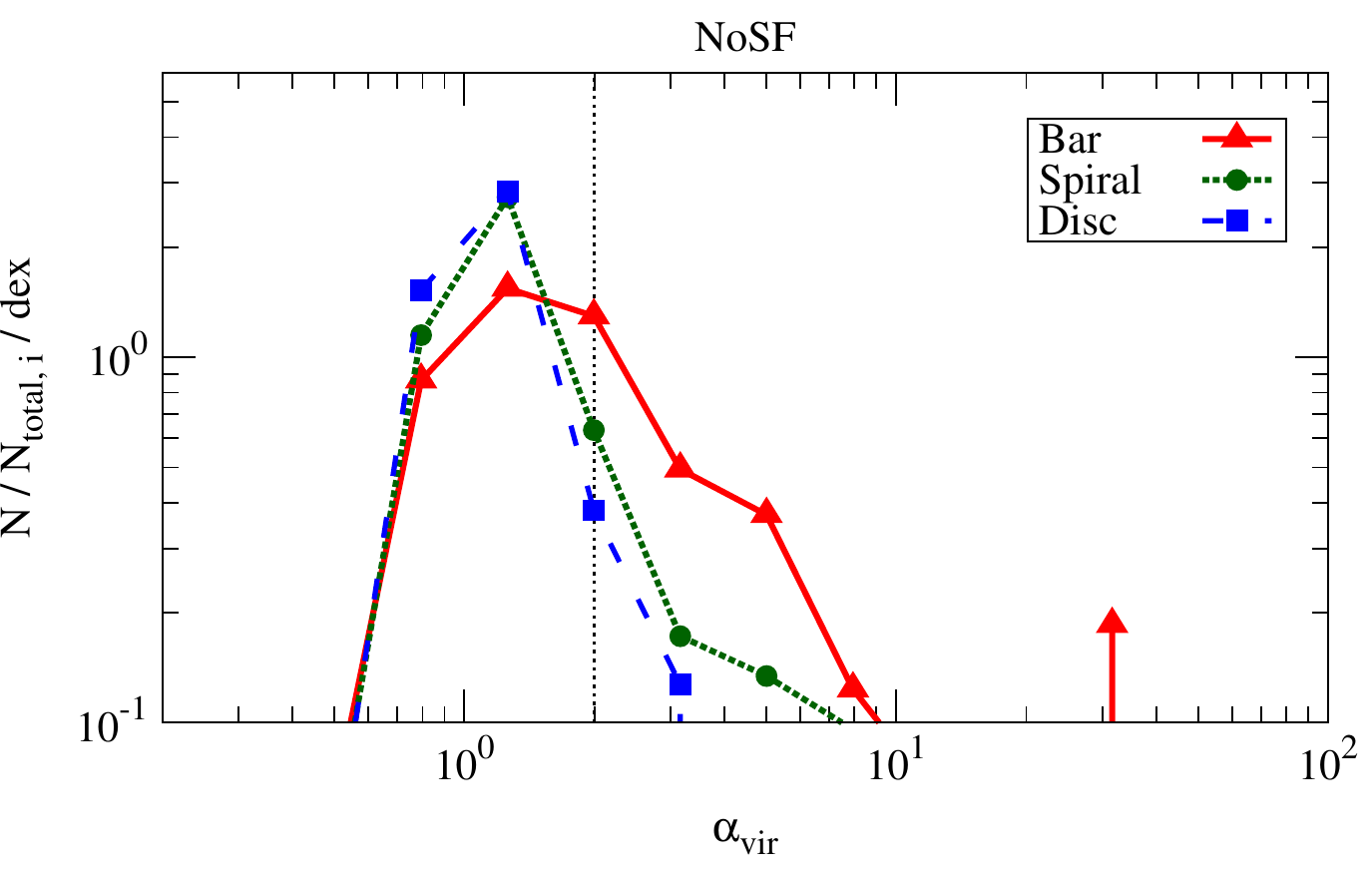}}
	\subfigure{
	\includegraphics[width=60mm, bb=0 0 396 252]{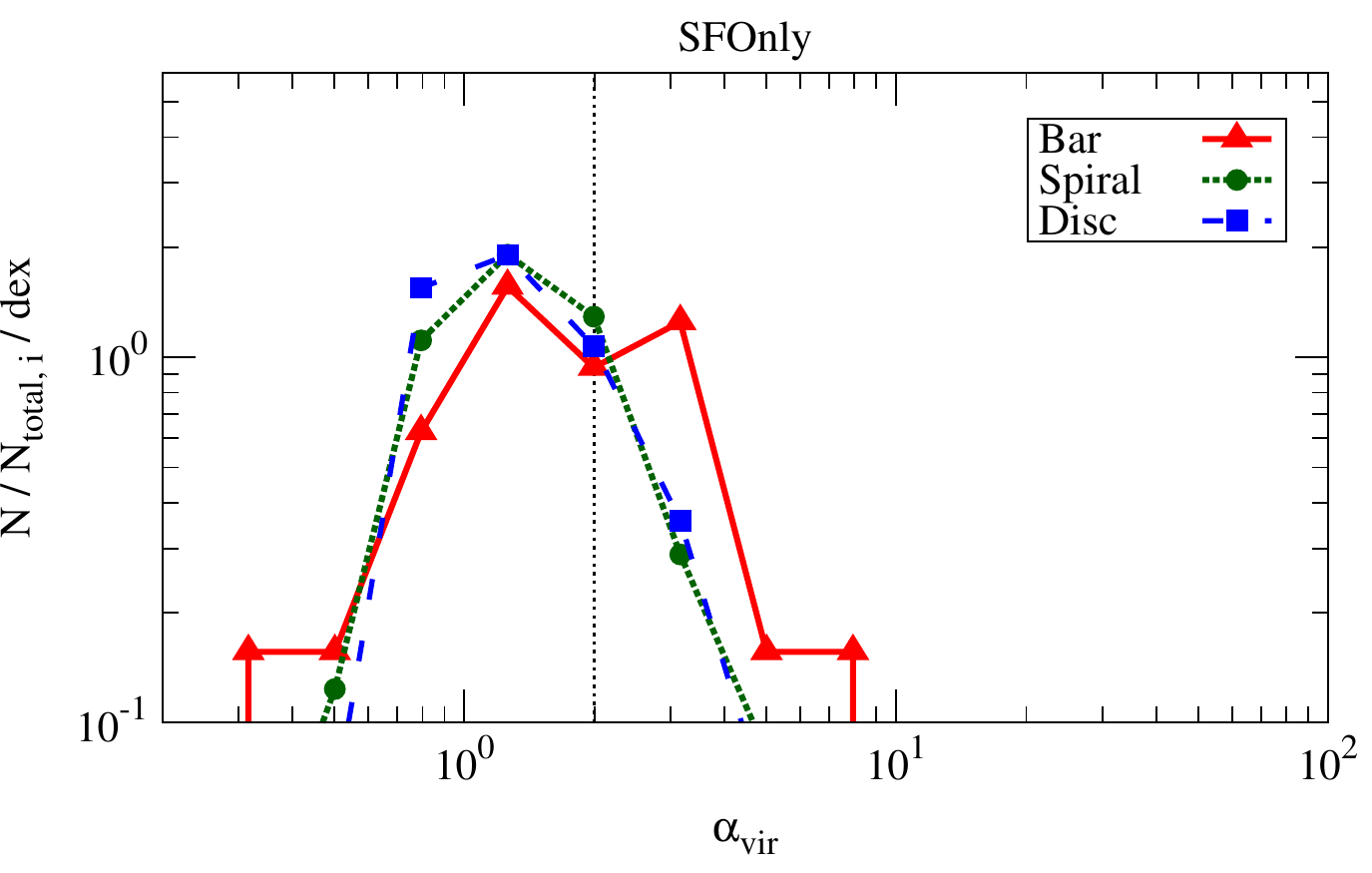}}
	\subfigure{
	\includegraphics[width=60mm, bb=0 0 396 252]{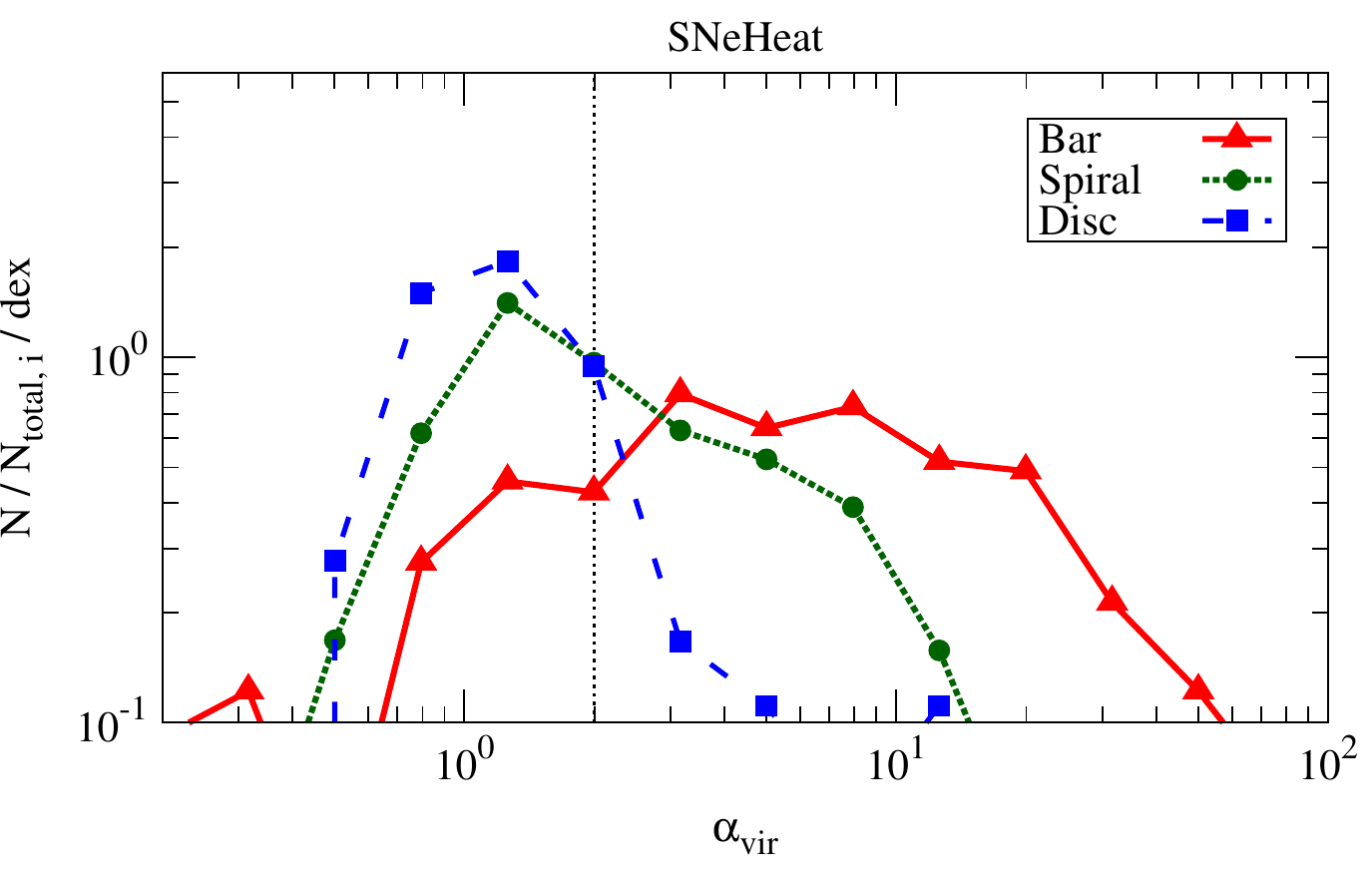}}
\end{tabular}
\caption{Normalised distributions of the cloud properties in the bar region (triangle solid lines), spiral region (circle dotted lines) and disc region (square dashed line) at $t =$ 200 Myr. From top to bottom, each row shows the cloud mass, the cloud surface density, $\Sigma_{\rm c} = M_{\rm c}/(\pi R_{\rm c}^2)$ and the virial parameter. From left to right are the cloud populations of the NoSF, SFOnly and SNeHeat runs.}
\label{distribution of cloud properties}
\end{center}
\end{figure*}

The distribution of properties within the cloud population itself is shown in Figure~\ref{distribution of cloud properties}. The top row of three panels shows the normalised distribution of cloud mass in the three runs. The population of clouds in the bar region (red solid line with triangular markers) is most strongly affected by the addition of stellar physics, while the lower density, quiescent outer disc region (Blue dashed line with square markers) is more uniform across all three runs. 

As star formation erodes the highest density gas, the maximum cloud mass is reined back from above $10^8$\,M$_\odot$ to about $3\times 10^7$\,M$_\odot$ in the bar region, where the highest fraction of {\it Type B} clouds form. This population of giant GMAs result in a distinctive bump in the bar profile that remains marked in all three simulations, peaking at around $10^7$\,M$_\odot$. The split between these {\it Type B} and the smaller {\it Type A} clouds occurs at around $M_{\mathrm{c}} =  5 \times 10^6\ \mathrm{M_{\odot}}$, corresponding to $R_{\mathrm{c}} = 30\ \mathrm{pc}$, where the border is marked in the scaling relations of Figure~\ref{scaling_relations}. The relative number of {\it Type B} clouds in the bar region matches that already seen in the bar chart in Figure~\ref{percentages_cloudtype_between_3regions}: the fraction rises in SFOnly and then drops in SNeHeat. This is not due to the absolute numbers of the GMAs increasing, but rather due to the fluctuations in the relative amount of the small {\it Type C} clouds. These can be seen at the low mass end of each distribution. In both the bar and spiral environments, the number of clouds below $M_c < 10^5$\,M$_\odot$ rises sharply when feedback is included in SNeHeat. By contrast, the disc region maintains a peak cloud value typical of that of the {\it Type A} clouds at $7\times 10^5$\,M$_\odot$, indicating that the outer parts of the disc are less influenced by feedback's effect on the surrounding ISM. 

The middle row of Figure~\ref{distribution of cloud properties} shows the surface density of the cloud populations. With the clear near parallel relations in Figure~\ref{scaling_relations} seen between the transient {\it Type C} and larger {\it Type A/B} clouds, these profiles are expected to be bimodal, with the dip at $\Sigma_{\mathrm{c}} = 230\ \mathrm{M_{\odot} pc^{-2}}$ that divides these two trends. This is seen in all simulations, but most prominently when feedback is included in SNeHeat, which bumps the {\it Type C} clouds to a population even larger than the {\it Type A} and {\it B} combined. A very clear trend can be seen going from disc to spiral to bar environments in the {\it Type C} low-mass mode, with the bar environment gaining the largest number of {\it Type C} clouds due to the influx of interstellar material from the outer parts of the disc. In contrast, the trend is weakest in the SFOnly run, where the formation of extremely massive clouds is throttled by star formation, reducing the filament environment where {\it Type C} clouds form. Without feedback to boost the filamentary structure in the warm ISM, the number of {\it Type C} clouds also declines.

The bottom row in Figure~\ref{distribution of cloud properties} shows the distribution of the cloud's virial parameters, with the vertical dotted line marking $\alpha_{\rm vir} = 2$, the boundary between gravitationally bound and unbound clouds. In NoSF, the clouds are predominantly bound (although only just) in all regions, with a peak at $\alpha_{\mathrm{vir}} \sim 1$. The bar region has a high-end tail of unbound clouds due to the high fraction of {\it Type C} clouds. Once star formation is included, this high-end unbound tail retreats, as the lower numbers of massive {\it Type B} clouds reduce the filament environment for {\it Type C} formation. The width of the distributions in SFOnly does increase, due to the boost in velocity dispersion from to the densest gas being removed to form stars, as discussed earlier. In SNeHeat, the drastic increase in {\it Type C} clouds greatly broadens the distribution, especially in the galaxy's central region. The peak in the bar environment moves from 1 to 3 to reflect the dominant {\it Type C} distribution.

%%%%%%%%%% subsection %%%%%%%%%%
\subsection{Cloud star formation}
\label{The effects of the warm ISM and type C clouds on star formation}

The previous section focussed on the changes in the cloud gas under different assumptions about star formation and feedback. This section considers the result of these variations on the cloud star formation. 

\begin{figure*}
\centering
	\includegraphics[width=10.0cm, bb=0 0 720 432]{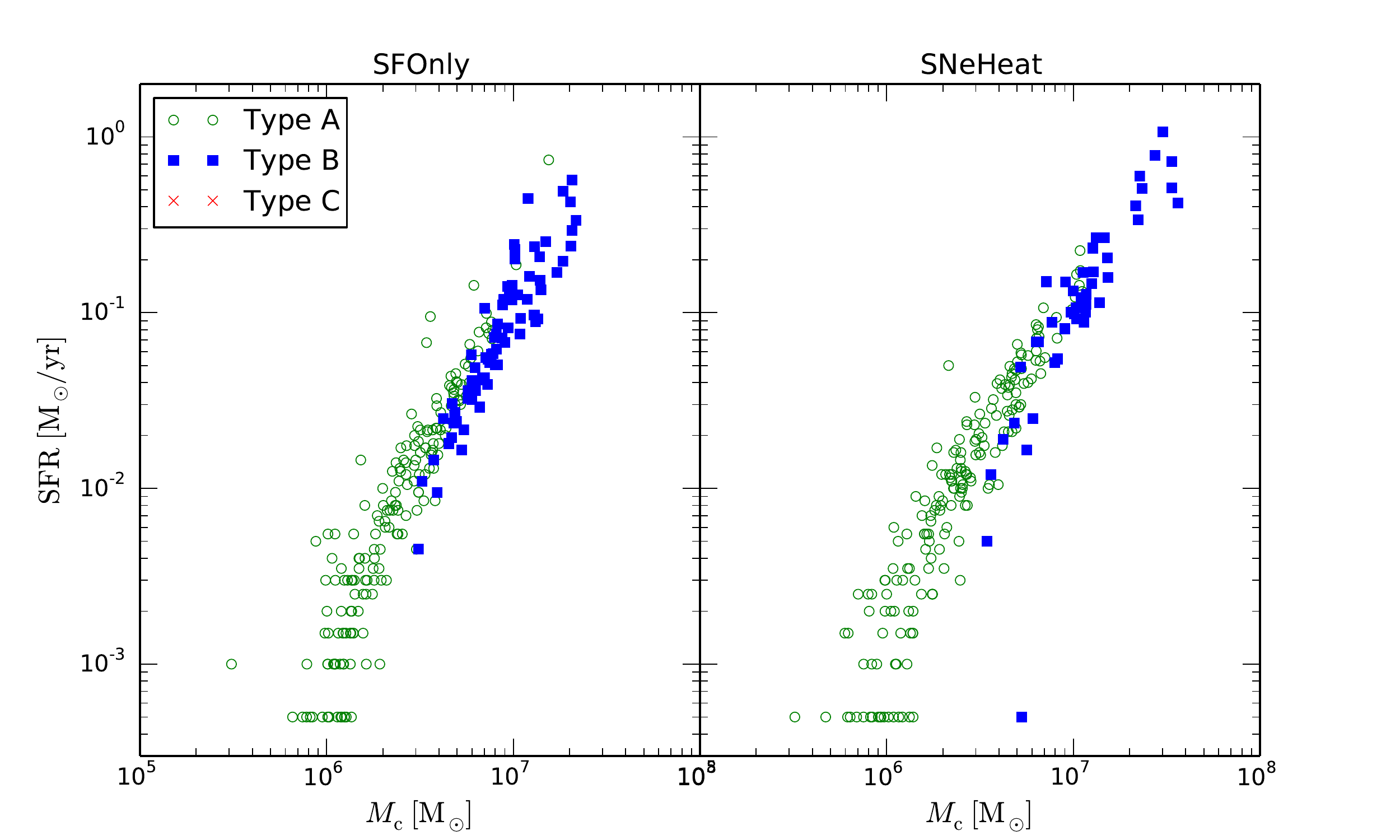}
	\caption{Cloud SFR (averaged over the previous 1\,Myr) versus cloud mass at $t =$ 200 Myr.  Type C clouds do not form stars and so do not show up in this plot.}
	\label{relationship_mass_stars}
\end{figure*}

The SFR per cloud as a function of its gas mass is shown in Figure~\ref{relationship_mass_stars} for the two simulations that include active star formation.  The estimate of the SFR within the cloud is calculated from the number of newly formed star particles within the preceding 1\,Myr that are born within the cloud boundary. 

The overall trend is similar for both SFOnly and SNeHeat, with a clear relationship between cloud gas mass and the resultant SFR. With their much larger mass, {\it Type B} clouds sit in the top-right of the relation, while the smaller {\it Type A} clouds are in the left-bottom. The transient {\it Type C} clouds cannot be seen on the plots at all, since their low surface densities and short lifetimes prevents them from reaching the densities needed to form star particles. 

Notably --unlike the mass-radius scaling relation in Figure~\ref{scaling_relations}-- the  {\it Type A} and {\it Type B} clouds do not quite form a continuation of the same trend. 
In the region around $3 \times 10^6 \sim 8 \times 10^6 M_{\odot}$, both {\it Type A} and {\it Type B} clouds overlap, but the {\it Type A} clouds have a consistently higher SFR for the same cloud gas mass. This difference is due to the large radii of the {\it Type B} clouds. Formed through multiple mergers of smaller clouds and defined as having radii above 30\,pc, the {\it Type B} population includes less compact objects that tend to feature an envelope of diffuse tidal interaction gas. By contrast, a {\it Type A} cloud of the same mass have a radius below 30\,pc, resulting in a denser cloud. This higher density gives {\it Type A} clouds a higher SFR, producing the observed offset.

When stellar feedback is included, the spread in the {\it Type A} clouds remains very similar, but the {\it Type B} clouds show variation at the low and high mass tails. In particular, at the high-mass end, the SNeHeat simulation contains a group of {\it Type B} clouds with significantly higher masses and SFRs than the other clouds in the simulation. These clouds were also seen in Figure~\ref{scaling_relations} and exist in the central bar region, which benefits most strongly from the gas inflow towards the galaxy centre. In this region, the gas density can get very high, with a corresponding high SFR. 

\subsection{The cloud drag force}

\begin{figure*}
\centering
	\subfigure{
	\includegraphics[width=8.0cm, bb=0 0 576 432]{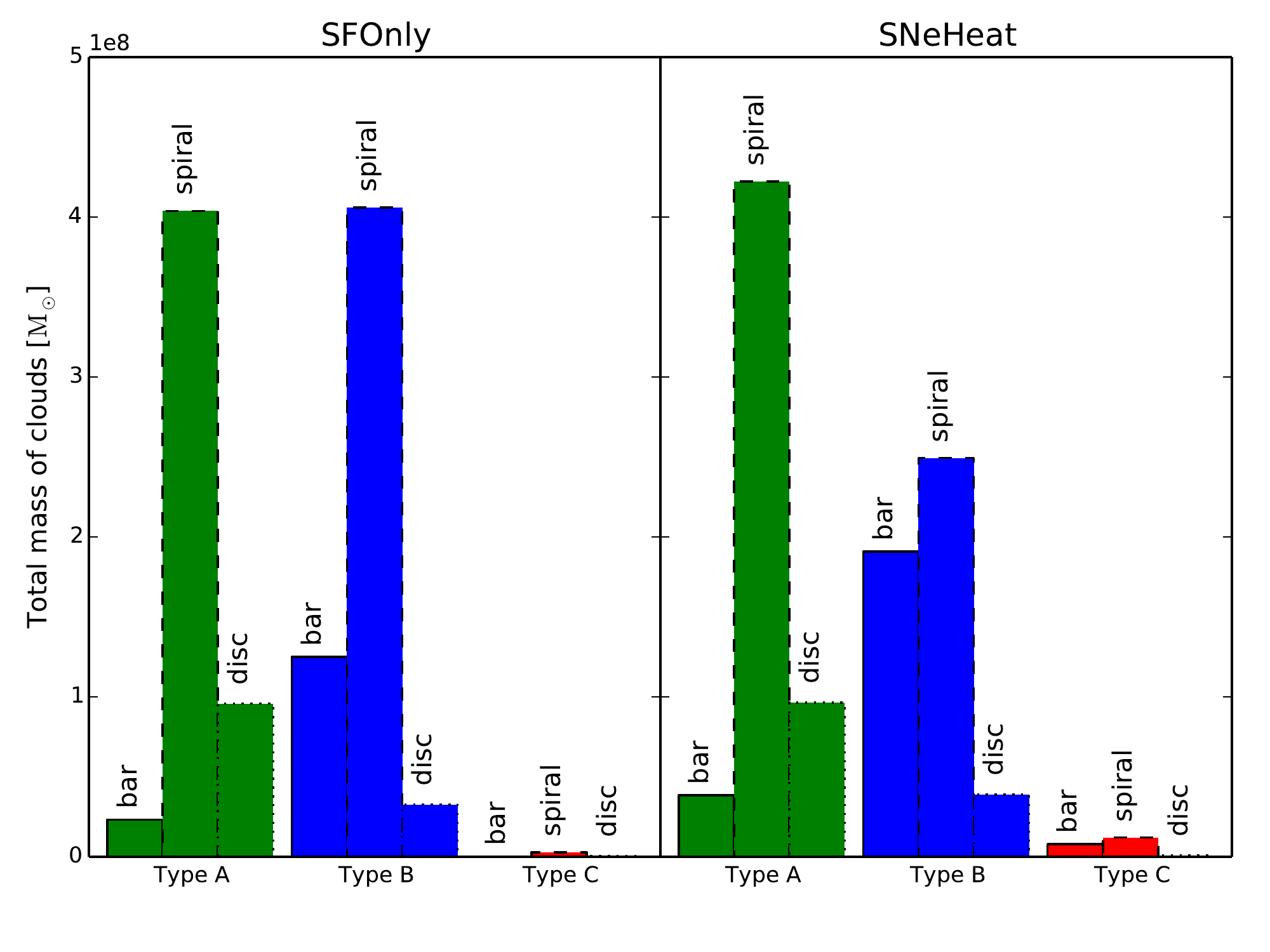}}
	\subfigure{
	\includegraphics[width=8.0cm, bb=0 0 576 432]{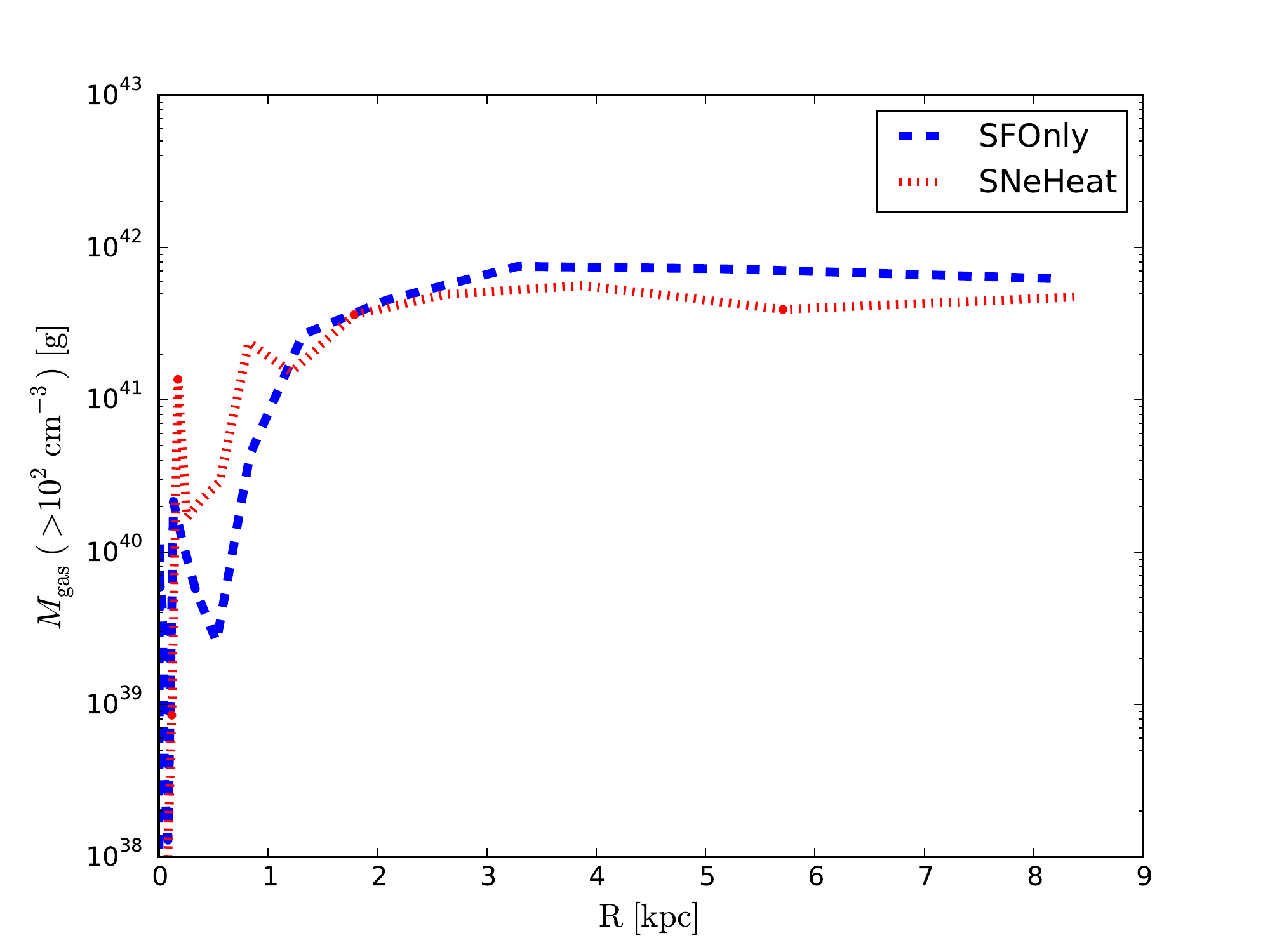}}
	\caption{Left: Total mass in clouds for the three different cloud types in each galactic region at $t = 200$\, Myr. Right: Radial distribution of the total gas mass denser than $100\ \rm cm^{-3}$ (our cloud threshold).}
	\label{total_mass_cloudtype_between_3regions}
\end{figure*}

As we have seen, the addition of stellar feedback boosts star formation in the galaxy's central regions by gas inflow; however, the mechanism for moving the gas is less obvious. This section takes a closer look at how gas flows may be driven.

First, we demonstrate that there is a net transfer of dense, star-forming clouds from larger to smaller radii.  The left-hand panel of Figure~\ref{total_mass_cloudtype_between_3regions} shows the total mass in clouds of each cloud type within the three galactic regions for the SFOnly and SNeHeat runs. While there is little difference in the {\it Type A} clouds, the amount of mass in {\it Type B} and {\it C} shows substantial change. 

For the massive (and strongly star-forming) {\it Type B} clouds, the amount of mass in the spiral region reduces and (largely) appears in the bar region as feedback is included. Note, this appears contrary to what was seen in Figure~\ref{numbers_cloudtype_between_3runs}, where the relative number of clouds dropped once feedback was included. That was due to the sheer number of forming transient {\it Type C} clouds, which contain very little mass. When mass is considered, the bar gets an obvious boost of {\it Type B} cloud material that moves inwards from the outer disc regions. Since cloud material ejected via thermal feedback would not result in a strong change in radii of cloud material, this boost in {\it Type B} cloud mass implies that the clouds themselves are moving inwards through the disc. The total amount of gas in {\it Type B} is slightly reduced compared to SFOnly, as outer layers are removed in feedback processes to convert some clouds into {\it Type A}s. However, this reduction is more than compensated for in the inner galaxy regions due to the substantial influx of cloud mass.

The right-hand panel in Figure~\ref{total_mass_cloudtype_between_3regions} shows the radial profile of the cloud-forming  gas (densities above 100\,cm$^{-3}$) for the SFOnly and SNeHeat runs. This shows the dense gas collecting in the central bar region ($R < 1\ \rm kpc$) when feedback is included, and a corresponding dip in the outer regions of the spiral and disc ($R > 3\ \rm kpc $). Cloud material is therefore moving from the outer regions to the inner regions, creating a boost in {\it Type B} cloud mass in the inner parts of the disc.

\begin{figure*}
\centering
	\includegraphics[width=8cm, bb=0 0 576 432]{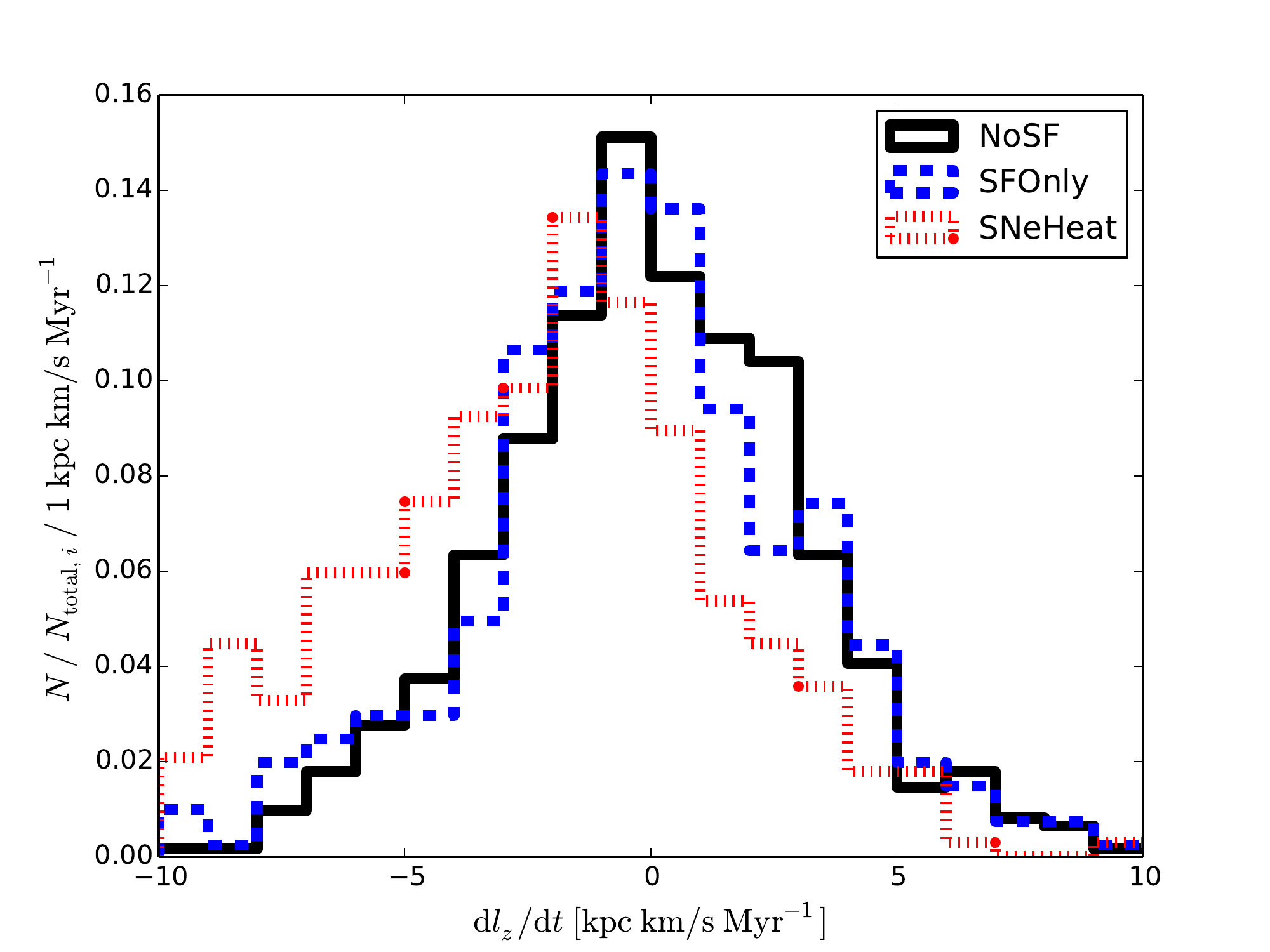}
	\caption{Normalised histogram of $\mathrm{d}l_z/\mathrm{d}t$, where $l_z$ is a $z$ component of the specific angular momentum of the clouds around the galactic centre. The variation is calculated between 199 and 201\,Myr as  $\mathrm{d}l_z/\mathrm{d}t = \{l_z(201\ \mathrm{Myr}) -  l_z(199\ \mathrm{Myr})\}/2\ \mathrm{Myr}$. We exclude the $type\ C$ clouds from this distribution due to their small size.}
	\label{distribution_dl_dt, distribution_dm_dt}
\end{figure*}

To confirm that clouds are actually moving through the disc, we analysed the angular momentum loss of the {\it Type A} and {\it Type B} clouds. The result is shown in Figure~\ref{distribution_dl_dt, distribution_dm_dt}, which plots the normalised histogram of the change in the $z$-component of the specific angular momentum, $l_z$, for each cloud around the galactic centre. To compute this change, we compared the specific angular momentum at 199 and 201\,Myr for each cloud,  $\mathrm{d}l_z/\mathrm{d}t = \{l_z(201\ \mathrm{Myr}) -  l_z(199\ \mathrm{Myr})\}/2\ \mathrm{Myr}$. {\it Type C} clouds are excluded in this distribution due to their small size and short lifetimes.

For the two runs without feedback, the angular momentum distribution peaks at $\mathrm{d} l_z / \mathrm{d} t = 0$, suggesting no net angular momentum change during the considered 2\,Myr. On the other hand, clouds in the SNeHeat simulation peak to the left, indicating that the clouds tend to lose angular momentum during their lifetime when stellar feedback is included. Such a loss will cause clouds to spiral in towards the galactic centre, as seen in Figure~\ref{total_mass_cloudtype_between_3regions}.

One possible mechanism for the movement of cloud material via angular momentum loss is the impact of drag. Drag opposes the cloud motion around the galaxy disc. The drag force, $\mathbf{F}_{\mathrm{drag}}$, is defined as,
\begin{align}
\mathbf{F}_{\mathrm{drag}} = \frac{1}{2} D \rho (\mathbf{v} - \mathbf{v}_{\mathrm{c}})^2 A\ \frac{\mathbf{v} - \mathbf{v}_{\mathrm{c}}}{|\mathbf{v} - \mathbf{v}_{\mathrm{c}}|},
\label{eq:drag}
\end{align}
where $D$ is a dimensionless drag coefficient of order unity, $\rho$ is the density of the surrounding ISM around the cloud, $\mathbf{v}$ is the velocity of the ISM, $\mathbf{v}_{\mathrm{c}}$ is the cloud velocity, and $A$ is a cross sectional area of the cloud. The surrounding density is calculated in the region the cloud is moving towards,  within a sphere that is twice the average radius of the cloud. This drag equation means that the denser the surrounding ISM, the stronger the drag.  Earlier, we say that stellar feedback resulted in an increase in the density of the intercloud medium.
This then would explain an increased drag force on the largest {\it Type B} clouds, pulling those objects towards the galaxy centre. Their motion would both explain the increase in {\it Type B} mass seen in Figure~\ref{total_mass_cloudtype_between_3regions} and the replenished gas supply to keep the star formation high. Drag is therefore a major effect from the addition of stellar feedback. 

This view can be further confirmed by estimating the drag force on each cloud and comparing this with the angular momentum loss. To estimate the magnitude of the drag force, the cloud cross sectional area is calculated as $A = \pi {R_{\rm c}}^2$. For rigid bodies, Newton's second law for rotation is
\begin{equation}
\mathbf{r} \times \mathbf{F}_{\mathrm{drag}} = \frac{\mathrm{d}}{\mathrm{d}t} \mathbf{L}
= m \frac{\mathrm{d}\mathbf{l}}{\mathrm{d}t} + \mathbf{l} \frac{\mathrm{d}m}{\mathrm{d}t},
\end{equation}
where $\mathbf{r}$ is a radial position of the cloud, $m$ is the cloud mass, $\mathbf{L} = m \mathbf{l}$ is the total cloud angular momentum, and $\mathbf{l}= \mathbf{r} \times \mathbf{v}$ is the specific angular momentum.

\begin{figure*}
\centering
	\subfigure{
	\includegraphics[width=8.0cm, bb=0 0 576 432]{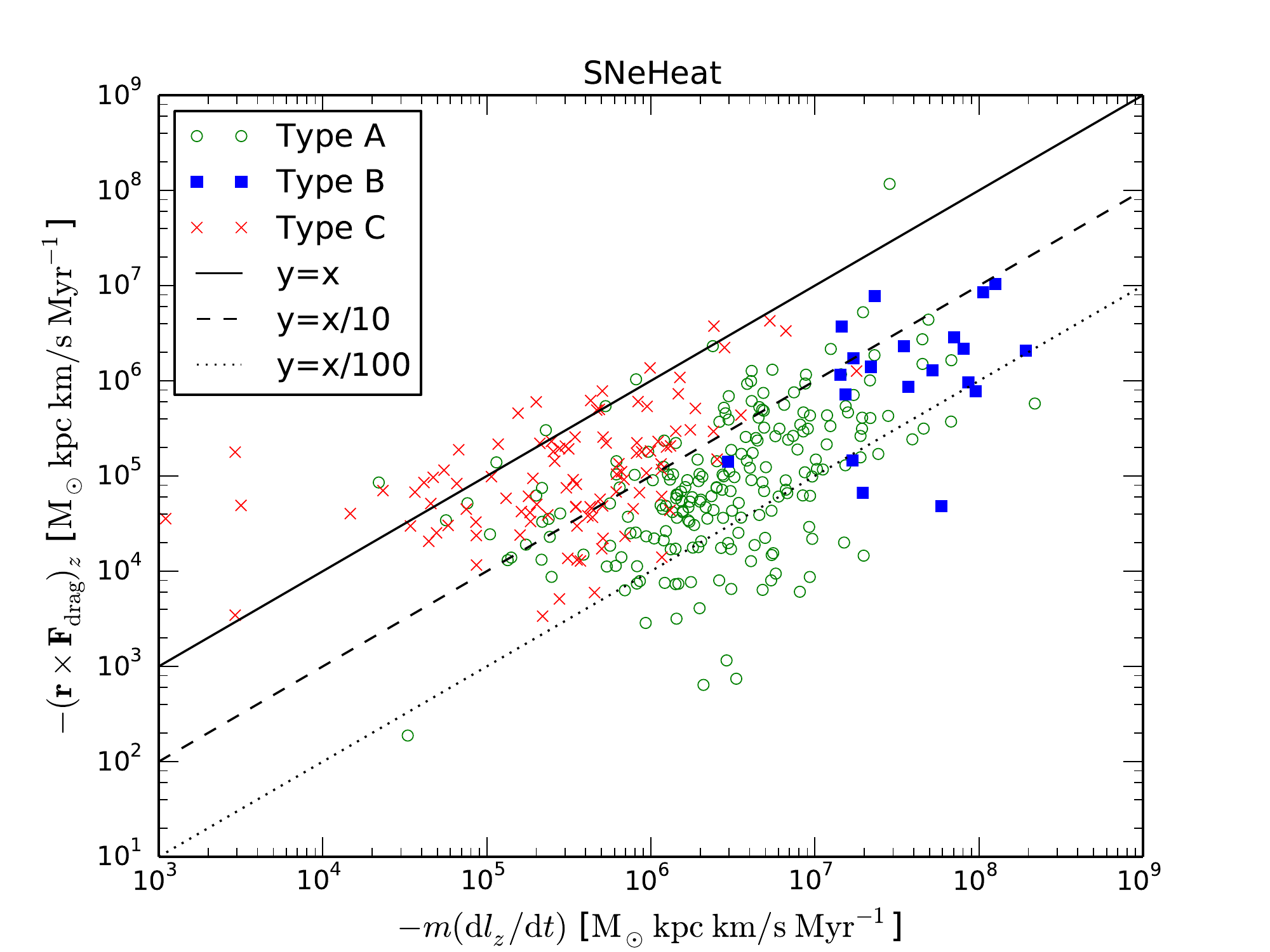}}
	\caption{Relation of $-(\mathbf{r} \times \mathbf{F}_{\mathrm{drag}})_z$ versus $-m(\mathrm{d}l_z/\mathrm{d}t)$ for each cloud. The green circles are the {\it Type A} clouds, the blue squares are the {\it Type B} clouds, and the red x are the {\it Type C} clouds.}
	\label{drag_force}
\end{figure*}

The relationship between the drag force and the loss in angular momentum can be seen in Figure~\ref{drag_force}. The graph shows $-(\mathbf{r} \times \mathbf{F}_{\mathrm{drag}})_z$ and $-m(\mathrm{d}l_z/\mathrm{d}t)$ (calculated to have the same units) for each cloud, ignoring the mass loss term, $l_z(\mathrm{d}m/\mathrm{d}t)$. Since both axes would contain negative values, the numbers are multiplied by $-1$ to make the scale positive. 

While the drag force term, $-(\mathbf{r} \times \mathbf{F}_{\mathrm{drag}})_z$, is about one order of magnitude less than the angular momentum loss term, $-m(\mathrm{d}l_z/\mathrm{d}t)$, there is a clear relation between them for the {\it Type A} and {\it Type B} clouds. The difference in magnitude means that factors in addition to drag must also be controlling the cloud motion. One of these is the mass loss from the cloud during the thermal energy injection. A normalised histogram (not shown for space) gave a value for the mass loss rate of the clouds, $dm/dt = \{M_{\rm c}(201 \ \mathrm{Myr}) - M_{\rm c}(199 \ \mathrm{Myr})\}/2 \ \mathrm{Myr} \sim -10^4$\,M$_\odot$Myr$^{-1}$ for SNeHeat run and $\sim 0$ for the NoSF and SFOnly runs. This mass loss when stellar feedback is included means that the $\mathbf{l} (\mathrm{d}m/\mathrm{d}t)$ must be calculated when estimating the drag force. The typical absolute value of the $l_z (\mathrm{d}m/\mathrm{d}t)$ is roughly the same order as that of $m (\mathrm{d}l_z/\mathrm{d}t)$.

A second factor is our simplified estimation of the drag force. The clouds are in reality poorly approximated as rigid bodies, since feedback will remove part of the gas and external forces distort the surface. Similarly, the cross-sectional area may be an underestimate, but this is hard to judge since clouds do not have a clear boundary \citep{Pan2015}. The exact value of the drag coefficient, $D$ is also unknown.

All these uncertainties mean that the magnitude of the drag force is a difficult quantity to pin down. However, the trend shown in Figure~\ref{drag_force} suggests that drag is playing a role in the reduction of the angular momentum of the larger clouds. For the small {\it Type C} clouds, there is no obvious relation between drag and angular momentum loss due to their small size. That is consistent with Figure~\ref{total_mass_cloudtype_between_3regions}, which showed that the main clouds being pulled inwards were the massive {\it Type B} clouds. 

Although previous works have investigated the effect of stellar feedback on GMCs, the process of the expelled gas creating a strong drag force on the clouds has previously been unreported. This might be due to the focus on the high density, star-forming gas in previous works, rather than the lower density warm ISM that plays a role here. 

Clouds can lose their angular momentum through additional differing physical processes. \citet{ZasovKhoperskov2015} showed that clouds could drop towards the galactic central region through dynamical friction between clouds, but they stated that lifetime of the clouds must be greater than 100 Myr; a time period inconsistent with observational estimates (5 $\sim$ 30 Myr. e.g. \citealt{Kawamura2009, Miura2012}) and our own results (typically less than 10 Myr).

\section{Discussion}

%%%%%%%%%% subsection %%%%%%%%%%
\subsection{Cooling and feedback}

How much damage stellar feedback inflicts on a GMC is hotly debated. In previous simulations, \citet{Tasker2015} found that the cloud is largely unaffected by thermal feedback, which exits the structure through the easiest route. Conversely, \citet{Williamson2014} finds that feedback plays a far more destructive role on the star formation nursery. 

\begin{figure*}
\centering
	\includegraphics[width=9.0cm, bb=0 0 576 432]{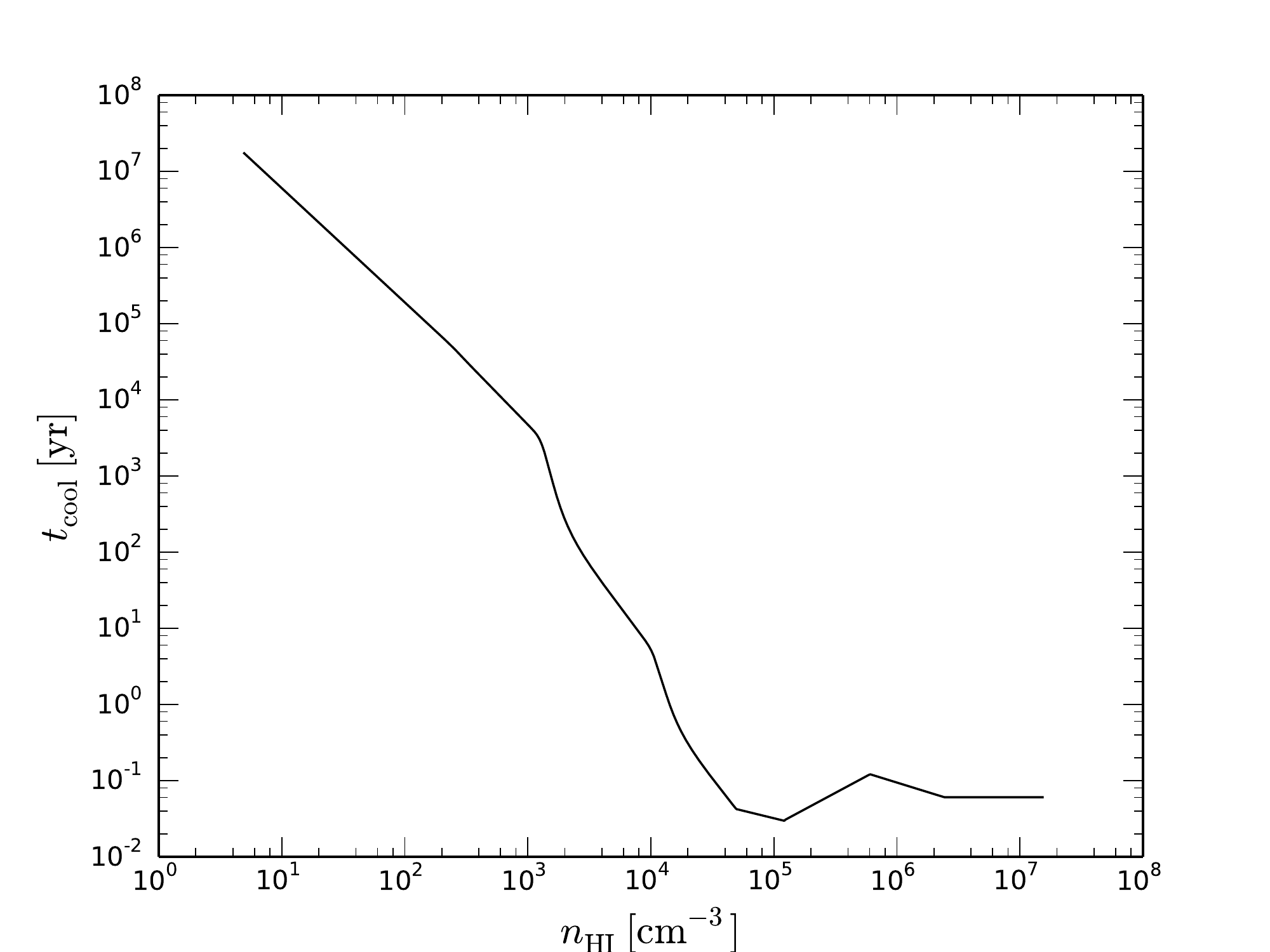}
	\caption{Cooling time of a cell injected with thermal energy from stellar feedback, depending on the cell gas density.}
	\label{density_coolingtime}
\end{figure*}

To assess the effectiveness of our feedback, we explored the time needed to radiatively cool away the energy injected during a supernovae event. The cooling time of the gas is given by the thermal energy divided by the cooling rate, $t_{\rm cool} = E/ \dot{E}$. Due to the magnitude of the thermal feedback, the energy is approximately equal to the injected energy. The maximum value for this in a cell is $E_{\rm cell} = f_{SN} m_{\rm star}c^2/19$, as described in Section~\ref{Star formation and feedback}, while the cooling rate is $\dot{E} = n_{\rm HI}^2\Lambda \Delta x^3$, for number density $n_{\rm HI}$ and cooling rate, $\Lambda$. Figure~\ref{density_coolingtime} shows the cooling time as a function of the number density. 

At our star formation threshold density at $10^4\ \mathrm{cm^{-3}}$, the cooling time is only a few years. This value is considerably shorter than the time step of the finest cell for the simulation ($dt \sim 200\ \mathrm{year}$). Thermal feedback injected into this region will cool rapidly and have no opportunity to affect the ISM. However, our model additionally adds thermal energy to the cells surrounding the star-forming cell. This region will contain lower density gas consistent with the cloud identification threshold of $100\ \mathrm{cm^{-3}}$. At these densities, the cooling time is about $10^5$ year, allowing time for the gas to respond.  This is consistent with the discussion in \citet{Simpson2015}.

Overcooling is a well-known problem in galaxy simulations (e.g. \citealt{Stinson2006}). Feedback energy injected into our densest cells will suffer from this issue, which makes the effects of our feedback a lower limit. Despite this, the result that the warm ISM can change the flow of star-forming gas in the simulation is unlikely to change.

%%%%%%%%%% Conclusion %%%%%%%%%%
\section{Conclusions}
\label{Conclusions}

We performed a simulation of an M83-type barred spiral galaxy with resolution down to the pc-scale, including star formation and thermal stellar feedback. By comparing simulations with different added stellar physics, we explored the effect of feedback on the properties of dense clouds in the bar, spiral and outer disc regions of the galaxy.  As initially described in Paper I, we divided our clouds into three types based on the mass-radius scaling relation: {\it Type A} clouds have properties typical to those of observed GMCs, {\it Type B} clouds were massive GMAs and {\it Type C} clouds were low density, transient objects. Our main results are as follows.

We find that, overall, the inclusion of star formation (without feedback) has only a minor impact on our simulated galaxy, primarily through the conversion of dense gas in large, self-gravitating clumps into stars.  In particular, we find that:

\begin{enumerate}

\item The addition of a star formation prescription allows us to estimate star formation rates on a cloud-by-cloud basis. The {\it Type B} clouds have the highest star formation rate, although their diffuse outer layers result in a slightly lower SFR compared to the more compact {\it Type A} clouds of similar mass. The diffuse {\it Type C} clouds do not produce any stars, as they do not become dense enough to reach the star formation threshold of $10^4$\,cm$^{-3}$.

\item The smaller number of massive clouds results in a decrease in the number of transient {\it Type C} clouds because these small clouds form out of the tidal debris produced from the large ({\it Type A} and {\it Type B}) clouds.

\end{enumerate}

The addition of feedback results in much more substantial changes due to outflows from the star-forming clumps and the resultant increase in the amount of gas in the interstellar medium.  Our primary results from the simulation including feedback are:

\begin{enumerate}
\setcounter{enumi}{2}

\item Stellar feedback does not usually destroy the GMCs, but it does disperse a substantial fraction of its gas mass. This gas flows into the inter-cloud region, raising the density of the warm ISM and reducing the mass in dense clouds.

\item The denser inter-cloud ISM becomes a new site for the formation of {\it Type C} transient clouds. Rather than forming in the dense filaments associated with tidal interactions, these new {\it Type C} clouds are formed in filamentary material in the inter-cloud ISM. This strongly increases their number compared to both the no-star formation and star formation only runs.

\item Massive clouds (especially {\it Type B} clouds) lose their specific angular momentum and move towards the inner galactic regions. The reason for this loss likely has multiple causes, but a trend between the drag induced by the higher density inter-cloud gas is one possible candidate.

\item The inflow of the dense gas towards the galactic centre supplies a significant amount of gas to the central bar region, replenishing that which is used during star formation.

\end{enumerate}

While further study will benefit from including additional forms of feedback such as radiation and ionising winds, this work strongly points at the importance of lower density inter-cloud gas. It is unlikely that star formation can be considered an entity controlled solely by the surround GMC nursery, instead being both affected and effecting the gas on a much wider scale.

\section*{Acknowledgments}
We acknowledge extensive use of the \textsc{yt} package \citep{Turk2011} in analyzing these results and the authors would like to thank the \textsc{yt} development team for their generous help.  Numerical computations were carried out on the Cray XC30 at the Center for Computational Astrophysics (CfCA) of the National Astronomical Observatory of Japan. YF is financially supported as Research Fellow of Japan Society for the Promotion of Science (JSPS KAKENHI Grant Number 15J02294). YF is funded by Clark Memorial Foundation, and Strategic Young Researches Overseas Visits Program for Accelerating Brain Circulation commissioned by the Japan Society for the Promotion of Science (R2405). EJT is funded by the MEXT grant for the Tenure Track System. GB acknowledges support from NSF grant AST-1312888, and NASA grant NNX15AB20G, as well as computational resources from NSF XSEDE, and Columbia University's Yeti cluster.  EJT and AH are supported by the Japanese Promotion of Science Sciety (KAKENHI Grant Number 15K05014). CMS acknowledges support from the European Research Council under ERC-StG grant EXAGAL-308037 and from the Klaus Tschira Foundation. Computations described in this work were performed using the publicly-available \textsc{Enzo} code (http://enzo-project.org), which is the product of a collaborative effort of many independent scientists from numerous institutions around the world. Their commitment to open science has helped make this work possible.

\bsp
\label{lastpage}
\end{document}